\documentclass[10pt,journal,compsoc]{IEEEtran}

\newcommand{\Duo}[1]{\textcolor{black}{#1}}
\newcommand{\tabincell}[2]{\begin{tabular}{@{}#1@{}}#2\end{tabular}}
\newcommand{\Revision}[1]{\textcolor{black}{#1}}

\ifCLASSOPTIONcompsoc
  \usepackage[nocompress]{cite}
\else
  \usepackage{cite}
\fi

\ifCLASSINFOpdf
\else
\fi

\usepackage{amsthm,amsmath,amssymb}
\usepackage{mathrsfs}
\usepackage{array}
\usepackage{graphicx}
\usepackage{epsfig}
\usepackage{subfigure}
\usepackage{xcolor}
\usepackage{booktabs}
\usepackage{multirow}
\usepackage{threeparttable}
\usepackage[linesnumbered,ruled,noend]{algorithm2e}

\begin{document}

\title{ILCAS: Imitation Learning-Based Configuration- Adaptive Streaming for Live Video Analytics with Cross-Camera Collaboration}

\author{Duo~Wu,
        Dayou~Zhang,~\IEEEmembership{Studuent~Member,~IEEE},
        Miao~Zhang,~\IEEEmembership{Studuent~Member,~IEEE},
        Ruoyu~Zhang,
        Fangxin~Wang,~\IEEEmembership{Member,~IEEE}
        and~Shuguang~Cui,~\IEEEmembership{Fellow,~IEEE}
\IEEEcompsocitemizethanks{\IEEEcompsocthanksitem Duo Wu, Dayou Zhang and Ruoyu Zhang are with the Future Network of Intelligence Institute and the School of Science and Engineering, The Chinese University of Hong Kong, Shenzhen, Shenzhen 518172, China.
Email: \{duowu,dayouzhang,ruoyuzhang\}@link.cuhk.edu.cn.
\IEEEcompsocthanksitem Fangxin Wang and Shuguang Cui are with the School of Science and Engineering and the Future Network of Intelligence Institute, The Chinese University of Hong Kong, Shenzhen and Guangdong Provincial Key Laboratory of Future Networks of Intelligence.
Email: \{wangfangxin,shuguangcui\}@cuhk.edu.cn.
\IEEEcompsocthanksitem Miao Zhang is with the School of Computing Science, Simon Fraser University, BC, Canada. Email: mza94@sfu.ca.
}
\thanks{Manuscript received xxx; revised xxx. 

\noindent (Corresponding author: Fangxin Wang)}}

\markboth{IEEE Transcations on Mobile Computing}%
{Wu \MakeLowercase{\textit{et al.}}: ILCAS: Imitation Learning-Based Configuration- Adaptive Streaming for Live Video Analytics with Cross-Camera Collaboration}

\IEEEtitleabstractindextext{
\begin{abstract}
The high-accuracy and resource-intensive deep neural networks (DNNs) have been widely adopted by live video analytics (VA), where camera videos are streamed over the network to resource-rich edge/cloud servers for DNN inference. Common video encoding configurations (e.g., resolution and frame rate) have been identified with significant impacts on striking the balance between bandwidth consumption and inference accuracy and therefore their adaption scheme has been a focus of optimization. 
However, previous profiling-based solutions suffer from high profiling cost, while existing deep reinforcement learning (DRL) based solutions may achieve poor performance due to the usage of fixed reward function for training the agent, which fails to craft the application goals in various scenarios. 
In this paper, we propose \texttt{ILCAS}, the first imitation learning (IL) based configuration-adaptive VA streaming system. Unlike DRL-based solutions, \texttt{ILCAS} trains the agent with demonstrations collected from the expert which is designed as an offline optimal policy that solves the configuration adaption problem through dynamic programming. To tackle the challenge of video content dynamics, \texttt{ILCAS} derives motion feature maps based on motion vectors which allow \texttt{ILCAS} to visually ``perceive'' video content changes. Moreover, \texttt{ILCAS} incorporates a cross-camera collaboration scheme to exploit the spatio-temporal correlations of cameras for more proper configuration selection. Extensive experiments confirm the superiority of \texttt{ILCAS} compared with state-of-the-art solutions, with $2$-$20.9\%$ improvement of mean accuracy and $19.9$-$85.3\%$ reduction of chunk upload lag.
\end{abstract}

}

\maketitle

\IEEEdisplaynontitleabstractindextext

\IEEEpeerreviewmaketitle

\ifCLASSOPTIONcompsoc
\IEEEraisesectionheading{\section{Introduction}\label{sec:introduction}}
\else
\section{Introduction}
\label{sec:introduction}
\fi

\IEEEPARstart{M}{illions} \Duo{of cameras today are widely deployed for live video analytics (VA) with various missions.  For instance, traffic cameras in modern cities need to detect vehicles and pedestrian for traffic control~\cite{abdullah2014traffic}. In disaster areas, unmanned aerial vehicles (UAVs) need to localize the survivors with their onboard cameras so that the survivors can be rescued as soon as possible~\cite{guo2022minimizing}. To guarantee high accuracy, numerous applications now employ deep neural networks (DNNs) for various analytics tasks, such as FasterRCNN for object detection~\cite{ren2015faster} and ICNet for semantic segmentation~\cite{zhao2018icnet}.} However, the current limited computation capacity at front-end devices prevents the computation-intensive DNN models from being conducted locally. One common method to overcome this limitation is to transmit the live videos from front-end cameras to resource-rich back-end servers to execute DNN inference for VA tasks~\cite{miao2023omnisense}.

In this context, the network bandwidth will become a bottleneck resource in live VA streaming due to the scarce bandwidth resources between cameras and servers~\cite{zhang2018awstream}.  Common video encoding configurations (i.e., combination of \textit{knobs} such as resolution and frame rate) have significant impacts on bandwidth consumption and inference accuracy in live VA streaming. For instance, an expensive configuration (e.g., high resolution and frame rate) with good video quality will ensure high accuracy but yield high bandwidth cost, which may entail long latency and analytic lag, and vice verse. 
Therefore, adjusting configurations to adapt to the variation of network bandwidth has been identified as a promising way to maximize the server-side DNN inference accuracy without exhausting bandwidth and causing analytic lag~\cite{zhang2018awstream}.

However, the design of such adaption strategy remains a significant challenge due to the highly dynamics and unstable network conditions. To make things worse, the optimal configurations also vary with the dynamic video contents. Intuitively, an ideal strategy should choose a low frame rate when the targets are moving slowly and a low resolution when they appear large in the visual field, since such choices will consume fewer network resources without affecting the accuracy. Yet practically, it is far more complex to model video content dynamics and capture the implicit relationship between configuration performance and video contents. 

Previous works usually use \textit{profiling-based} strategy to seek the best configurations, i.e., searching the possible combination of knobs on a small portion of a given video clip and applying the derived best one to the rest portion~\cite{zhang2017live}\cite{jiang2018chameleon}.  
For example, AWStream~\cite{zhang2018awstream} exhaustively profiles all configurations on a video clip to derive the profile that characterizes the bandwidth and accuracy trade-off of each configuration. It then online  adjusts the video configurations based on such profile and consistently updates the profile to maintain its freshness.
The key limitations of these works lie in two aspects. First, the profiling process requires performing analytics with \textit{golden} configuration (uses the best and most expensive values for all knobs)~\cite{zhang2017live}\cite{hsieh2018focus} on a part of video to obtain accuracy, while such \textit{golden} configuration will consume much bandwidth and entail long latency. 
\Duo{Second, the profiling-based strategies actually has an implicit assumption that the configuration selected through profiling will keep ``best'' for a certain time window, which in fact cannot adapt to the dynamics of video contents or work well under unexpected network conditions.}

\Duo{Recent \textit{learning-based} solutions~\cite{zhang2022casva}\cite{zhang2022maxim} use deep reinforcement learning (DRL) to learn an agent for selecting configurations without the streaming of golden configuration or the profiling process, which outperform traditional profiling-based solutions. 
They manually define a reward function to quantify the trade-off between VA accuracy and upload lag and uses such function to train their agent. The trained agent is used to directly select configurations based on the observed environment states (e.g., estimated network bandwidth).
Nevertheless, these solutions still suffer from three key problems. 
First, the performance of their DRL agent may drift far from the optimal because of the usage of fixed reward function. Specifically, they introduce fixed weight parameters in the reward function to characterize the trade-off between different performance metrics, while in practice different applications or scenarios have different emphasis of these metrics. 
Moreover, they only use coarse-grained numerical features to capture video dynamics, and therefore lack effectiveness to adapt to the changes of video contents. 
Additionally, they do not explore the spatio-temporal correlations of closely-located cameras to achieve more robust performance.
}

\Duo{In this paper, to combat the limitations above, we propose \texttt{ILCAS}, a novel learning-based framework for configuration-adaptive streaming for live VA, which aims to maximize the VA task accuracy while minimizing the upload lag for each chunk. The heart of \texttt{ILCAS} is an imitation learning (IL)~\cite{hussein2017imitation} method to train the agent. Unlike previous learning-based solutions that use handcrafted rewards for training, \texttt{ILCAS} trains the agent with a set of demonstrations collected from the expert, which is an offline optimal policy that solves the adaptive configuration selection problem with an efficient dynamic programming algorithm. With IL, the agent learns to solve the task by observing and imitating the expert's behavior from demonstrations, instead of explicitly optimizing a reward function. This enables \texttt{ILCAS} to achieve better trade-off between VA accuracy and upload lag, and therefore remarkably improves its overall performance.}

\Duo{Besides, to tackle the challenge of video dynamics, \texttt{ILCAS} extracts motion vectors from the video encoding process, and derives \textit{motion feature maps} that provide fine-grained information of content changes. With such features, \texttt{ILCAS} is able to ``perceive'' video dynamics and quickly adapt to content changes. In addition, the increasingly dense deployment of camera networks offers an opportunity for cross-camera collaboration to achieve more propoer VA configuration adaption. This is because cameras deployed in the same regions or along the same roads often share some spatial and temporal correlations. Hence, \texttt{ILCAS} also discovers the camera correlations and leverages a cross-camera collaboration scheme to select more accurate configurations and adapt to more complex scenarios.}

\Duo{
To summarize, the main contributions of this paper are as follows:
\begin{itemize}
    \item We propose \texttt{ILCAS}, the first IL-based framework for configuration optimization for live VA, which eliminates the weaknesses of handcrafted rewards that commonly exist in previous DRL-based methods.
    \item We propose a lightweight and useful indicator \textit{motion feature map} to infer video dynamics, which enables \texttt{ILCAS} to quickly adapt to the content changes of video.
    \item We explore the spatio-temporal correlations in a camera network and design an efficient cross-camera collaboration scheme, which allows \texttt{ILCAS} to achieve more robust performance in complex scenes. 
    \item We evaluate the performance of \texttt{ILCAS} over various real-world videos and network traces using trace-driven simulations. Results demonstrate that compared to state-of-the-art profiling-based and DRL-based solutions, \texttt{ILCAS} improves the mean accuracy by $15.0$--$20.9\%$ and $2.0$--$9.6\%$, while reducing the mean lag by $57.4$--$85.3\%$ and $19.9$--$74.0\%$, respectively. In addition, we also verify the effectiveness of \texttt{ILCAS} in the multi-camera environment.
\end{itemize}}

\Duo{The rest of this paper is organized as follows. Section~\ref{sec:motivation} introduces the motivation of this work. Section~\ref{sec:system_design} provides the system overview of \texttt{ILCAS}.  Next, the methodology of extracting motion feature map is presented in Section~\ref{sec:motion_feature_map}. We elaborate the detailed design of IL-enabled adaptive configuration selection in Section~\ref{sec:il_based_configuration_selection} and present the cross-camera collaboration scheme in Section~\ref{sec:cross_camera_collaboration}. We then conduct extensive experiments to evaluate the performance of \texttt{ILCAS} in Section~\ref{sec:Evaluation}. The related work is provided in Section~\ref{sec:related_work}. Finally, Section~\ref{sec:conclusion} concludes this paper.}

\section{Motivation}
\label{sec:motivation}

\subsection{Inferring Video Dynamics with Motion Feature Map}

\begin{figure}[t]
    \centering
    \subfigure[Frames of a video chunk]{
    \includegraphics[width=0.48\textwidth]{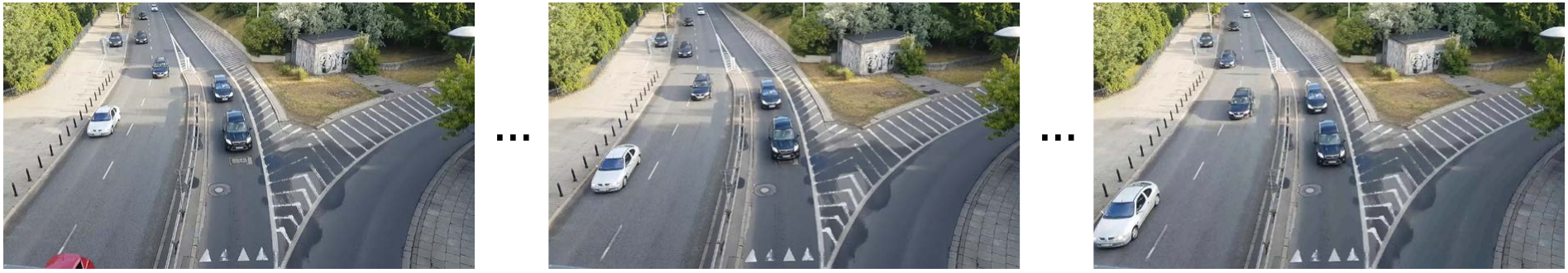}
    \label{subfig:mv_visualization}
    }
    \subfigure[Extracted motion feature map]{
    \includegraphics[width=0.25\textwidth]{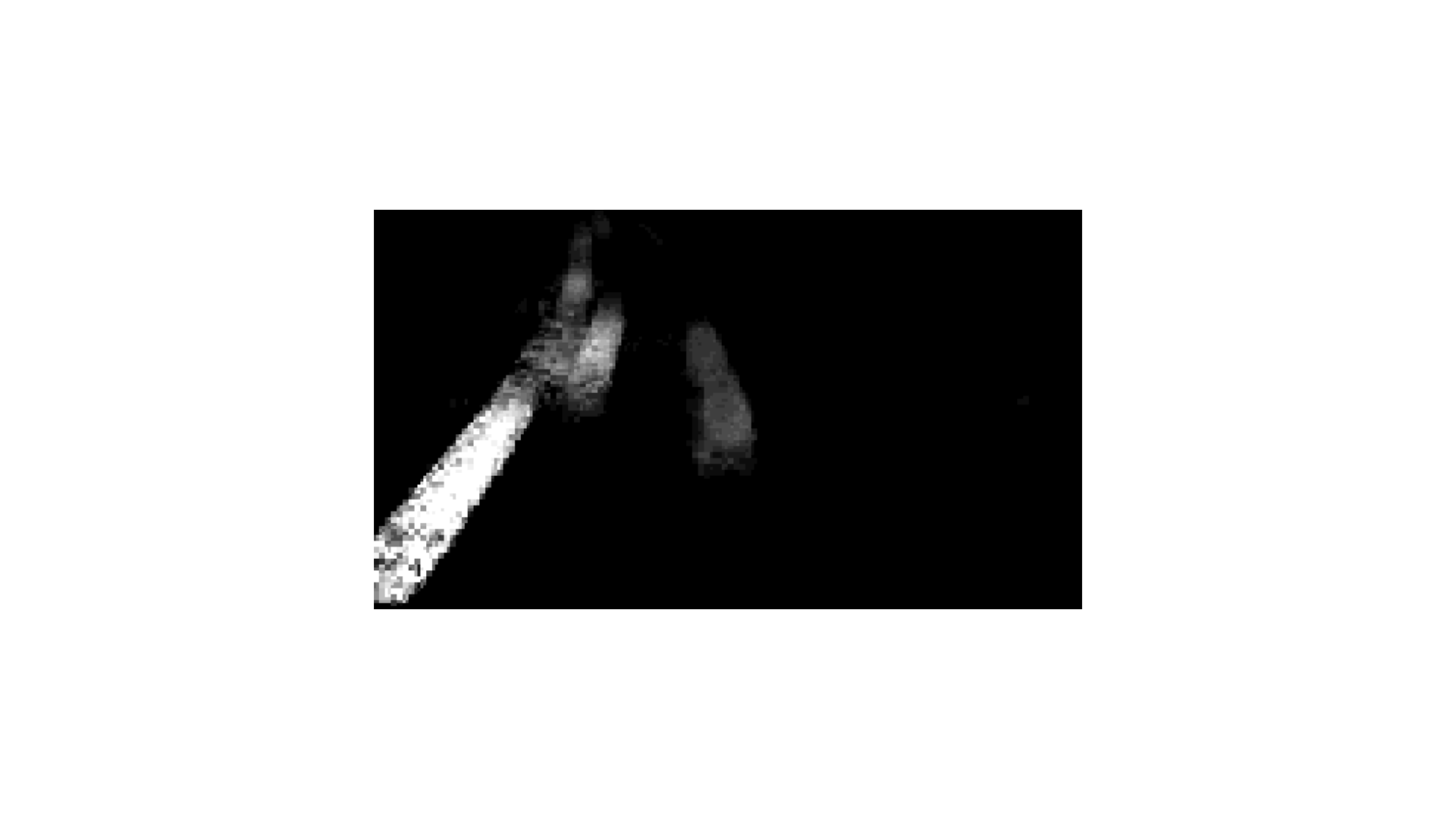}
    \label{subfig:mv_visualization}
    }
    
    \caption{Illustration of motion feature map. The brighter a region is, the more significant content changes in that region are.}
    \label{fig:motion_feature_map_vis}
\end{figure}

One fundamental challenge of VA configuration optimization is how to handle the fine-grained video content dynamics.
The optimal configuration varies with dynamic video content, and their relationship is hard to model or predict.
\Duo{One way to infer video dynamics is using optical flow, a concept widely studied in computer vision community. Despite its capability to provide pixel-level information of content changes, the calculation of optical flow is time-consuming~\cite{Zhang_2016_CVPR}, thus preventing its usage in live VA. Existing work~\cite{zhang2022casva} proposed to quantify video dynamics by comparing the volumes of current and previous chunks in the same configuration. The rationale behind is that an increasing bitrate means increasing content dynamics, which may cause performance degradation of a configuration. Yet, this approach only provides coarse-grained information and introduces additional encoding overhead~\cite{zhang2022casva}.}

\Duo{Our goal is to find a lightweight indicator that can accurately capture the fine-grained video content dynamics. To achieve this, we resort to motion vectors, the products in common video codecs such as MPEG-2 and H.264.}
In an encoded video stream, each frame is divided into macroblocks with pre-determined size, which ranges from $4\times4$ to $16\times16$ pixels. Video encoder adopts a block matching algorithm to find a block similar to the one that is encoded on the reference frame. 
\Duo{The absolute positions of source and destination of a macroblock then define a motion vector $\vec{m}= ((src_x, src_y ), (dst_x, dst_y ))$. By definition, motion vectors naturally capture the content changes of frames, and thus open new opportunities for inferring video dynamics. Hence, in this paper, we use motion vectors to derive lightweight and useful \textit{motion feature maps} to characterize content dynamics. We extract motion vectors from the encoding process and calculate the motion feature maps on the camera, then send these feature maps as grayscale images to the configuration controller to assist it making decisions (see Section~\ref{sec:motion_feature_map}). Figure~\ref{fig:motion_feature_map_vis} illustrates the visualization of a motion feature map.
As shown, if the region is brighter, the content changes in that region are more significant.
It is clear that motion feature maps intuitively describe the content changes of video, allowing \texttt{ILCAS} to visually ``perceive'' the video dynamics and make more adaptive configuration decisions. Besides, as shown in Table~\ref{tab:motion_feature_map_overhead}, the computation and transmission overhead of motion feature maps is negligible. For instance, for a 1080p video, the calculation of a feature map only takes 47ms, far less than the chunk length of 1s. Moreover, its size (3.47KB) accounts for less than $1\%$ of the chunk size\footnote{Note that the motion feature map is calculated at chunk-level, not frame-level.}.  
}

\begin{table}[t]
    \centering
    \caption{Overhead of Motion Feature Maps. }
    \label{tab:motion_feature_map_overhead}
    \vspace{-0.2cm}
    \begin{threeparttable}
    \begin{tabular}{m{1.65cm}<{\centering} m{1.65cm}<{\centering}  m{1.65cm}<{\centering}  m{1.65cm}<{\centering} }
        \toprule
        \multirow{2}{*}{Resolution\tnote{$\dagger$}}  & \multirow{2}{*}{\tabincell{c}{Computation \\Time\tnote{$\ddagger$}}} & \multirow{2}{*}{Image Size}  & Image Size\\
        \cline{4-4}
        & & & Chunk Size \\
        \midrule
        240p & 2ms & 0.47KB & 0.98\%\\ 
        \midrule
        480p & 9ms & 1.47KB & 0.79\%\\  
        \midrule
        720p & 20ms & 2.47KB & 0.64\%\\  
        \midrule
        1080p & 47ms & 3.47KB & 0.31\%\\  
        \bottomrule
    \end{tabular}
     \begin{tablenotes}
        \footnotesize
        \item[$\dagger$] The frame rate and quantization parameters of each video are fixed to 30 and 29, respectively. The chunk length is 1 second.
        \item[$\ddagger$] Time measurement is performed on an Intel i5-8265U CPU. Task process is restricted to use only 1 CPU core.
      \end{tablenotes}
    \end{threeparttable}
\end{table}

\subsection{From Profiling-based to Learning-based Adaptation}

\begin{figure}[t]
\centering
\subfigure[Profiling-based strategy (coarse-grained)]{
\includegraphics[width=8.cm]{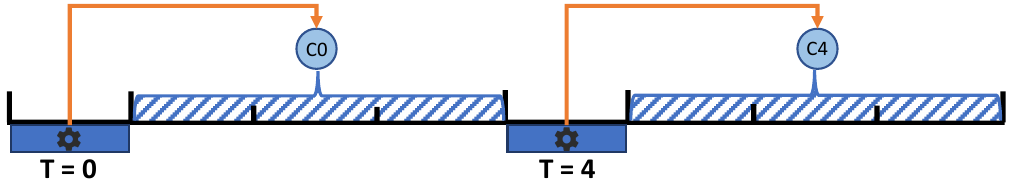}
\label{profiling_based}
}
\subfigure[Learning-based strategy (fine-grained)]{
\includegraphics[width=8.cm]{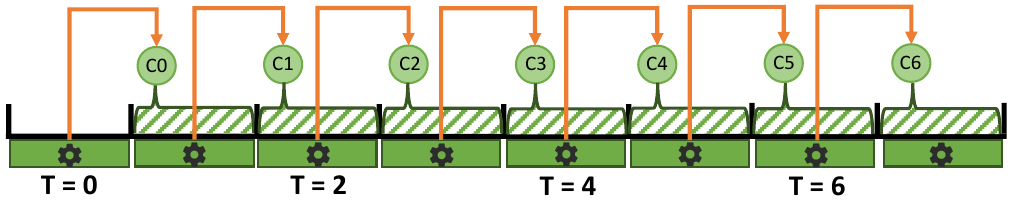}
\label{learning-based}
}
\caption{Comparison of profiling-based and learning-based strategies.}
\label{grain}
\end{figure}

Previous works mostly utilize the \textit{profiling-based} strategy to seek the best configuration. 
As illustrated in Figure~\ref{profiling_based}, they extract a small portion of video (e.g., $T_0$) streamed in \textit{golden} configuration to profile the ``best'' configuration.
The ``best'' configuration is then applied to the rest of the video clip (e.g., $T_1$ to $T_3$). This strategy, however, suffers from two key limitations. First, each profiling process requires to stream a golden part as the ground truth to compute accuracy.
This strategy indeed wastes scarce bandwidth as this profiling window usually accounts for 40\% to 75\% of the overall resources consumption through our empirical observation of previous work~\cite{jiang2018chameleon}. 
Second, since profiling is expensive, the gap between each profiling window has to be large enough (usually 4 seconds~\cite{jiang2018chameleon}) to reduce the profiling cost, which therefore only achieves \textit{coarse-grained} adaptation and fails to adapt to the dynamics of video contents. 

To address this problem, we propose the \textit{learning-based} strategy, as shown in Figure~\ref{learning-based}, which achieves \textit{fine-grained} adaptation (e.g., 1 second in our later experiments). The key idea is an offline training phase where the model learns from past videos, and an online inference phase using the well-trained model without streaming golden configuration, which saves substantial bandwidth resources. Since the inference phase has little overhead, the analytics window can be small enough to achieve fine-grained adaption.

\subsection{From Reinforcement Learning to Imitation Learning}
\Duo{
The heavy reliance of reinforcement learning (RL) on well-defined reward functions restricts its applications in complex tasks, where crafting such functions is extremely difficult~\cite{le2022survey}.
In the context of configuration-adaptive video streaming, it is non-trivial to quantify the trade-off between VA accuracy and upload lag or resource consumption. The reason lies in the complex relationships between different performance metrics as well as the varying preferences of high-level applications for them. In consequence, RL may lead to poor performance or uncontrolled behavior.}

\begin{figure}[t]
    \centering
    \subfigure[Mean accuracy]{
    \begin{minipage}[b]{0.2\textwidth}
    \includegraphics[width=1\textwidth]{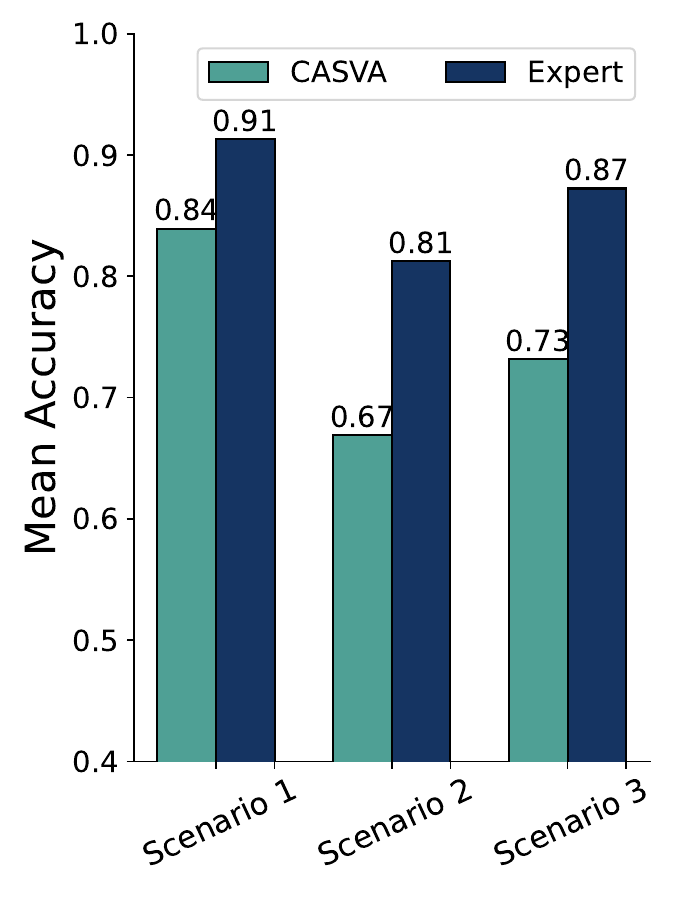}
    \label{subfig:mean_acc_gap}
    \vspace{-0.5cm}
    \end{minipage}
    }
    \subfigure[Mean lag]{
    \begin{minipage}[b]{0.2\textwidth}
    \includegraphics[width=1\textwidth]{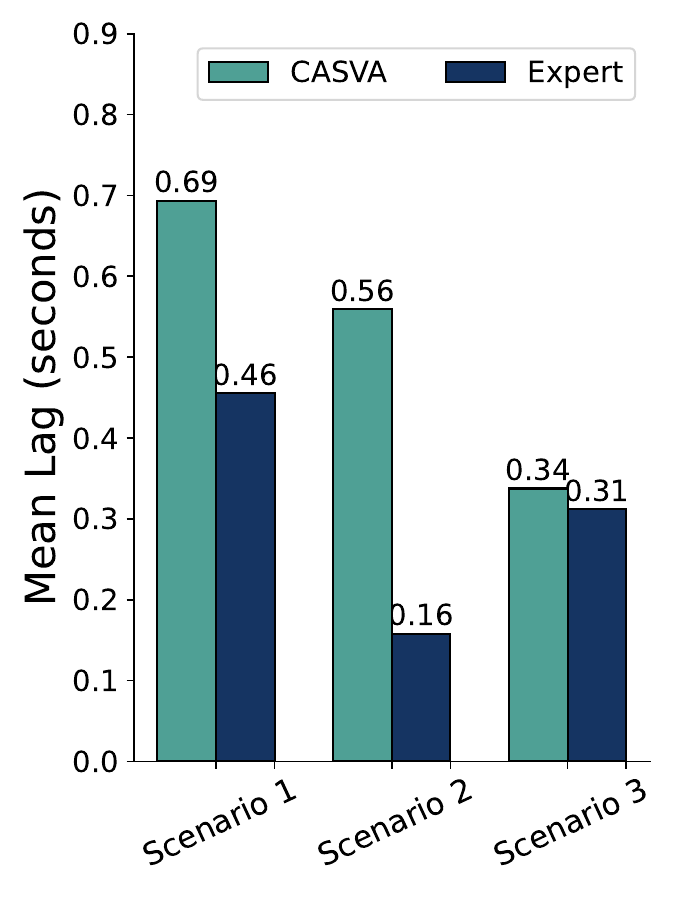}
    \label{subfig:mean_lag_gap}
    \vspace{-0.5cm}
    \end{minipage}
    }
    \caption{Illustration of the ineffectiveness of training agent with handcrafted rewards (Scenario 1: stationary video; Scenario 2: dash video; Scenario 3: stationary video and poor network condition).}
    \label{fig:explicit_rewards}
\end{figure}

\Duo{Existing DRL-based solutions~\cite{zhang2022casva,zhang2022maxim} explicitly define reward as a linear function with fixed weight parameters to measure the trade-off between different metrics. Such function is then used to train their agent. 
Although this approach may produce acceptable results in some cases, the agent's performance may drift far from the optimal especially in some scenarios where the fixed reward function fails to characterize the trade-off. 
We take CASVA~\cite{zhang2022casva}, the state-of-the-art DRL-based solution, as an example to illustrate the ineffectiveness of this approach. We vary the camera videos and network traces while fixing the weight parameters of reward function in CASVA, and use CASVA to select configurations in difference scenarios. We consider object detection as the vision task, and accuracy and upload lag as performance metrics. Following previous work~\cite{jiang2018chameleon}, the detection accuracy of a video chunk is calculated as the fraction of frames with F1 scores $\ge 70\%$, while F1 scores are computed by comparing the detected bounding boxes of golden configuration and selected configuration. The Intersection over Union (IoU) threshold for counting true positive detection is set to $50\%$. The upload lag of a chunk is defined as the difference between its expected upload time and actual upload time~\cite{zhang2022casva}. A larger lag means longer latency, thus decreasing the efficiency of live VA.}

The measurement results are reported in Figure~\ref{fig:explicit_rewards}. It can be seen that in Scenario 1, although CASVA produces $50\%$ more lag than the expert (an offline optimal policy, see Section~\ref{sec:il_based_configuration_selection}), its mean accuracy is only $7\%$ lower than the expert. 
However, when the video type changes (Scenario 1 $\rightarrow$ Scenario 2), the overall performance gap between CASVA and expert significantly increases due to the higher complexity (e.g., faster scene changes) and larger bitrate of dash video. Besides, when the network condition becomes poor (Scenario 1 $\rightarrow$ Scenario 3), CASVA aggressively selects cheap configurations to ensure small lag in sacrifice of VA accuracy, which therefore results in a larger accuracy gap ($7\% \rightarrow 14\%$). 
The poor performance of CASVA stems from the usage of fixed reward function, which fails to guide the agent to properly select configurations to balance the inference accuracy and upload lag.
To conclude, learning by reward function fails to achieve the best trade-off and suffers from the scalability issue. While it is possible to improve the performance of CASVA in Scenario 2/3 by re-tuning the weight parameters of its reward function, such fine-tuning process, in practice, is manually performed in a trial-and-error manner~\cite{sutton2018reinforcement}, which is labor-intensive and time-consuming.

In this paper, we leverage the advanced imitation learning (IL)~\cite{hussein2017imitation} technique to train our agent from expert's demonstrations. We design an expert to teach the agent how to solve the task of configuration selection. In this context, the agent learns the expert's underlying pattern and imitates the expert's behavior to select configurations, which allows our agent to achieve better trade-off and improve its overall performance. Moreover, thanks to the expert's guidance to solve the task in different scenarios, the scalability of our agent is also greatly improved.

\subsection{From Single-camera to Cross-camera Adaption}
\label{sec:cross_camera_enhancement}

\begin{figure*}[t]
    \flushleft 
    \begin{minipage}[t]{0.2\textwidth}
    \centering
    \includegraphics[width=\textwidth]{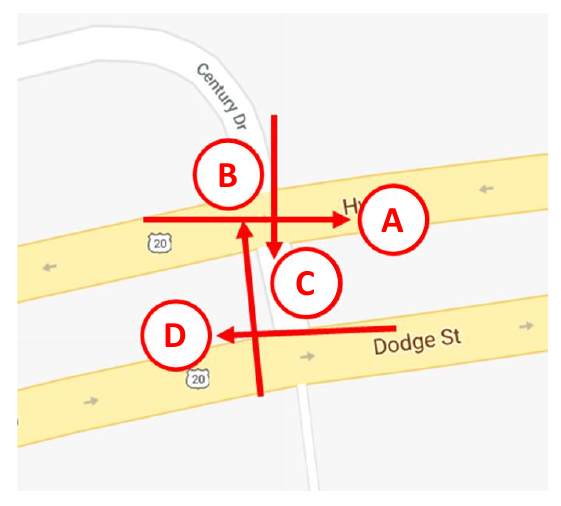}
    \vspace{-0.5cm}
    \vspace{-0.3cm}
    \caption{Topology of a real-world camera network~\cite{cross}.}
    \label{fig:cross_location}
    \end{minipage}
    \hspace{0.3cm}
    \begin{minipage}[t]{0.75\textwidth}
    \centering
    \includegraphics[width=\textwidth]{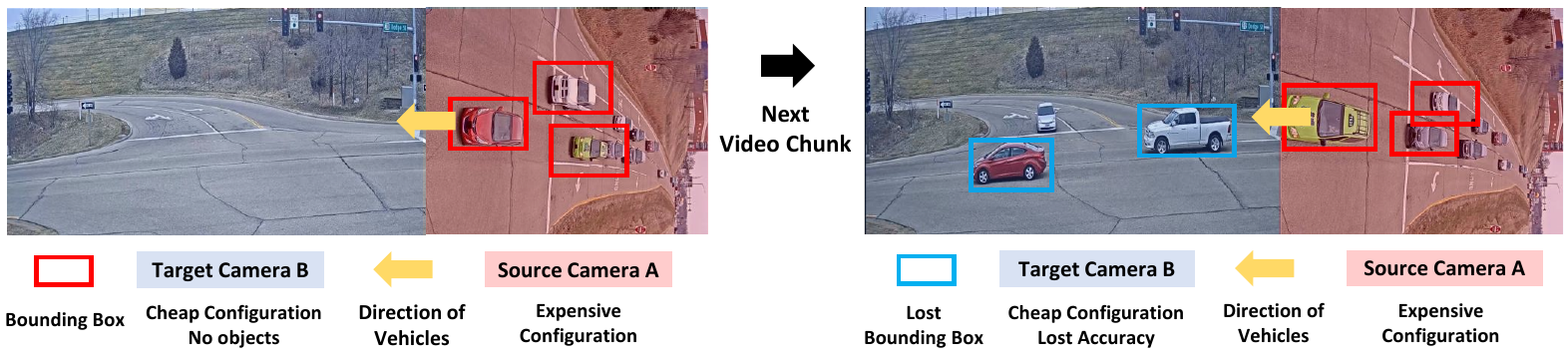}
    \vspace{-0.5cm}
    \caption{Limitation of the single-camera-based adaption.}
    \label{fig:single-limi}
    \end{minipage}
\end{figure*}

\begin{figure}[t]
    \centering
    \subfigure[Spatial Correlations]{
    \begin{minipage}[b]{0.23\textwidth}
    \includegraphics[width=\textwidth]{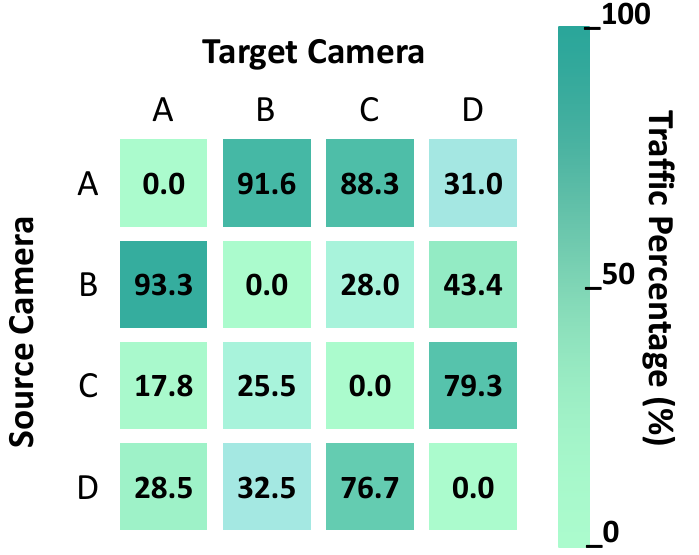}
    \label{subfig:cross_spatial}
    \vspace{-0.5cm}
    \end{minipage}
    }
    \subfigure[Temporal Correlations]{
    \begin{minipage}[b]{0.23\textwidth}
    \includegraphics[width=\textwidth]{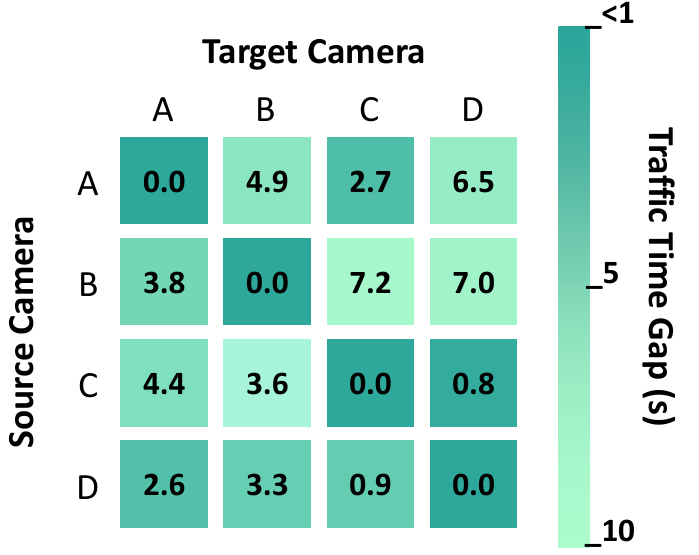}
    \label{subfig:cross_temporal}
    \vspace{-0.5cm}
    \end{minipage}
    }
    \caption{The spatial and temporal correlations of cameras. }
    \label{fig:cross_correlations}
\end{figure}

In a camera network, each camera can only acquire partial information due to the limited field of view (FOV), which restricts the performance of single-camera-based configuration adaption. Take camera pair A and B in Figure~\ref{fig:cross_location} as an example, whose FOVs are illustrated in Figure~\ref{fig:single-limi}. In this scenario, camera B applies a cheap configuration (e.g., low frame rate and resolution) to save network resources since there is no object in the view. However, the information in camera A's FOV indicates that many vehicles will appear in camera B's FOV in the near future. Since camera B is unaware of the upcoming of vehicles due to the limited FOV, it will still keep a cheap configuration in the following video chunk and thus fails to detect those vehicles, resulting in accuracy loss. This problem can be addressed if we exploit the camera relationships for configuration adaption. For instance, the accuracy loss of camera B can be avoided if it is alarmed of the upcoming vehicles with the FOV information from camera A.

Nevertheless, utilizing the camera relationships for more accurate configuration adaption is challenging, due to the complex spatio-temporal correlations of cameras. Figure~\ref{fig:cross_correlations} shows the camera correlations of the camera network in Figure~\ref{fig:cross_location}. Note that the sum of each row and column in Figure~\ref{subfig:cross_spatial} exceeds 100\% since there is overlap between cameras' FOV, i.e., the same object may appear in different cameras' FOV within a certain time. As shown, 88.3\% of vehicles from camera A will next be detected in camera C in 2.7 seconds on average. In that case, camera A can be the source camera to target camera C, so camera C can utilize the information from camera A for configuration adaption. However, it is not feasible to take camera C as the source camera, and camera A as the target camera in turn since only 17.8\% of vehicles appear in A's FOV after appearing in C's FOV. The reason is that most of the objects appearing in camera C's FOV next move to camera D's FOV, which indicates that specific analysis and quantification is necessary for utilizing the spatio-temporal information of cameras. 

To tackle the above challenges, in our cross-camera collaboration scheme, we quantify the spatio-temporal correlations of cameras and introduce an \textit{info-sharing} mechanism to share the FOV information with motion feature maps across cameras. Leveraging the motion feature maps from correlated cameras, each camera can infer the future video dynamics more accurately, thus achieving more proper configuration adaption.

\section{System Overview}
\label{sec:system_design}

\begin{figure*}[t]
  \centering\centerline{\epsfig{figure=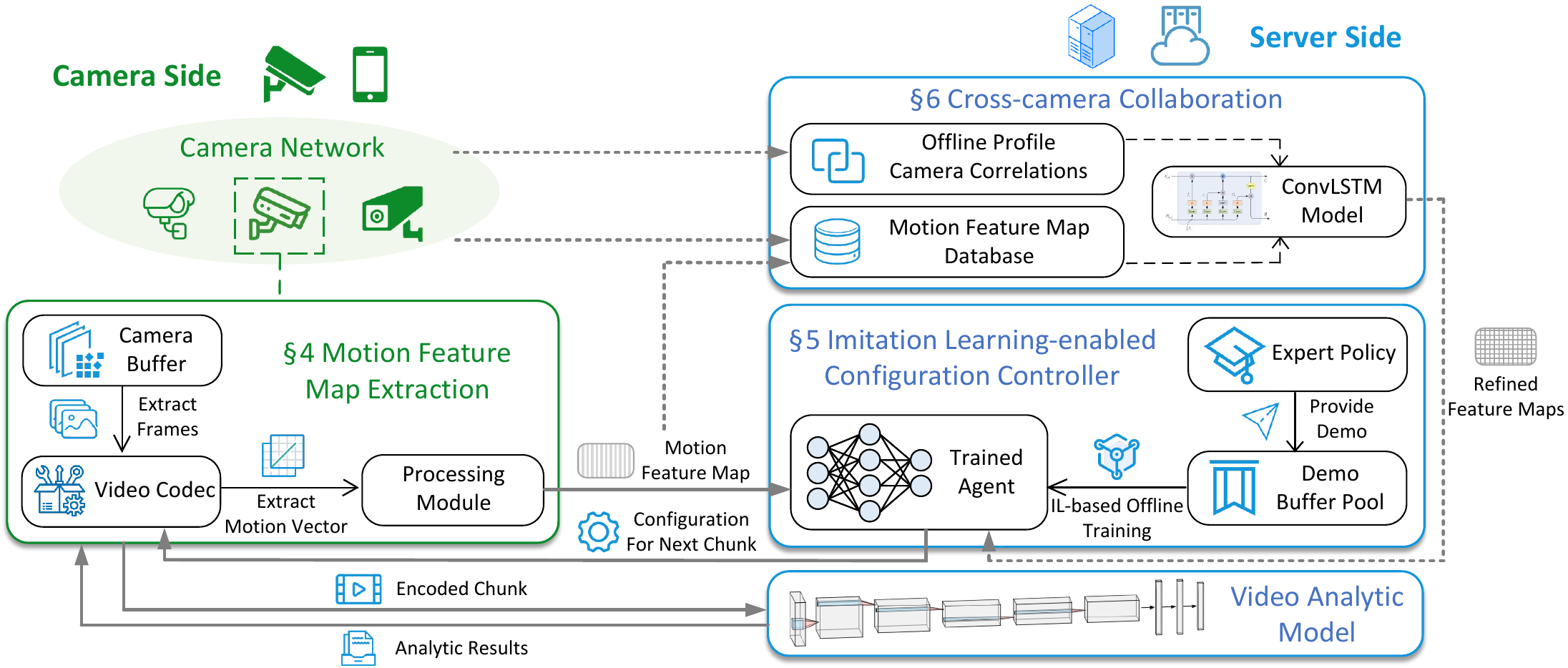,width=0.92\textwidth}}
  \caption{System framework of \texttt{ILCAS}.}
  \label{fig:system}
\end{figure*}

\Duo{Figure~\ref{fig:system} demonstrates the system overview of \texttt{ILCAS}, which consists of three core components: motion feature map extraction module, IL-enabled configuration controller and cross-camera collaboration scheme.}

\Duo{\textit{Motion feature map extraction}. On the camera side, the camera extracts raw frames from its buffer, encodes them with specific configuration, and sends the encoded chunk to the server for DNN inference (e.g., Faster RCNN~\cite{ren2015faster}, Yolo~\cite{redmon2016you}). During the encoding process, the camera extracts motion vectors from its codec, and computes the motion feature map which reflects the content changes of video. Such feature map is then sent to the configuration controller to assist the controller selecting configurations.}

\Duo{\textit{IL-enabled configuration controller}. Upon receiving the motion feature map, the controller observes the environment state (e.g., network throughput, camera buffer size) and uses its well-trained agent to select the configuration for encoding the next chunk. The agent is offline trained with the advanced IL framework, where an expert policy is designed to provide high-quality demonstrations for guidance. Considering the limited computing power of front-end devices, we implement the configuration controller on the resource-abundant edge/cloud servers.}

\Duo{\textit{Cross-camera collaboration (optional)}. When the target camera is deployed in a dense camera network, \texttt{ILCAS} can also exploit the information of other correlated cameras to achieve more robust configuration adaption. Specifically, \texttt{ILCAS} identifies the spatio-temporal correlations of cameras offline, collects and stores their motion feature maps in the database, and leverages an efficient Convolutional Long Short-Term Memory (ConvLSTM)~\cite{ConvLSTM} model to extract hidden features from these motion feature maps. The outputs of ConvLSTM model are then fed into the agent for selecting more appropriate configurations for the target camera.}

\section{Motion Feature Map Extraction}
\label{sec:motion_feature_map}

\Duo{
As the product of video encoding, motion vector naturally crafts the content changes of video. 
Specifically, motion vector describes the positions of two similar macroblocks in the encoded and reference frame. For instance, let $(x, y)$ represent the center of a block in the current encoded frame, and $(x',y')$ be the center of its similar block in the reference frame, then vector $\vec{m}=((x',y'),(x,y))$ defines a motion vector. For a block with center $(x,y)$ and motion vector $\vec{m}=((x',y'), (x,y))$, we define its \textit{motion degree} as the Manhattan Distance of its motion vector, so as to quantify its movement information:
\begin{equation}
    m_d(x,y)=|x-x'|+|y-y'|
\end{equation}}

\Duo{As there are multiple encoded and reference frame pairs within a chunk, a macroblock may be associated with multiple motion vectors. Hence, we accumulate all the motion degrees of a block $(x,y)$, and obtain its overall motion degrees within the chunk:
\begin{equation}
    M_d(x,y)=\sum m_d(x,y)
\end{equation}}

\Duo{In this way, we can obtain the matrix $M_d$ recording the overall motion degrees of all blocks. To derive the motion feature map, we further apply the clipping and scaling functions on matrix $M_d$~\cite{Zhang_2016_CVPR}. Specifically, each element of $M_d$ is clipped by:
\begin{equation}
    M_d^{clip}(x,y)=clip(M_d(x,y) / f, 0, 255)
\end{equation}
where $f$ is the frame rate of the encoded chunk. Next, by scaling $M_d^{clip}$ with the following function, the motion feature map $M$ of the current chunk is obtained:
\begin{equation}
    M(x,y) =\left\{
        \begin{aligned}
        255 & , & \text{if} \ M_d^{clip}(x,y) \ge \sigma, \\
        \frac{255}{\sigma} M_d^{clip}(x,y)  & , & \text{otherwise}
        \end{aligned}
        \right.
\end{equation}
where we empirically set $\sigma=20$. 
}

\Duo{
Note that in some codecs such as H.264, the macroblocks may have variable sizes (e.g., $4 \times 8$, $16 \times 16$ pixels), which prevents us from directly summing up the motion degrees of blocks with the same center position. To tackle this problem, we divide each macroblock into several microblocks with a fixed size of $4 \times 4$ pixels\footnote{The size of a microblock can be set to other values, depending on specific codecs.} and assign the motion degree of such macroblock to its corresponding microblocks. 
For example, an $8 \times 8$ macroblock is divided into 4 microblocks. Assume the motion degree of this block is 10, then each of its microblock is assigned with motion degree 10.
In this context, the motion degree of a microblock can be represented by one pixel value, and thus the size of the resulting motion feature map is reduced by $4\times4=16$ times. 
This can also significantly reduce the transmission overhead and the controller's computation overhead of processing the feature map.
}

\Revision{The extracted motion feature maps provide microblock-level information of video content changes. The size of a microblock is $4\times 4$ pixels, which accounts for 0.02\% of a 240p video with $320\times240$ pixels. Therefore, even in low resolution videos, motion feature maps can still capture fine-grained changes of video contents, enabling the configuration controller to perceive video dynamics and select more appropriate configurations. Furthermore, the computation and transmission overhead of a motion feature map is also negligible. According to the measurements reported in Table~\ref{tab:motion_feature_map_overhead}, for a 1080p video, the calculation of a feature map only takes 47ms, and its size is only 3.47KB. It's worth noting that motion feature map is extracted at the chunk level (i.e., one feature map per chunk) rather than frame level. This means that the motion feature maps can be utilized to capture fine-grained video content dynamics with minimal overhead.}

\section{IL-based Adaptive Configuration Selection}
\label{sec:il_based_configuration_selection}
\Duo{In this section, we present the detailed design of IL-based adaptive configuration selection module in \texttt{ILCAS}. We begin by describing the optimization problem of configuration selection in live VA, then we propose an efficient IL-based method to solve the problem.}

\subsection{Problem Description}
\Duo{In a live VA pipeline, frames are consistently captured by the camera and encoded as video chunks to transmit to the server for analytics. Assume that the duration of each chunk is $T$ seconds. The VA system works on a series of chunks $\{1, 2, ..., N\}$, where $N$ can be infinite for continuous analytics. 
Let $\boldsymbol{c}_i \in \mathcal{C}$ denote the configuration applied on chunk $i$ in the encoding process, where $\mathcal{C}$ is the set of all possible configurations. In this paper, we consider a configuration as the combination of three knobs, i.e., $\boldsymbol{c}_i = \{r_i, f_i, q_i\}$, where $r_i, f_i, q_i$ represent the resolution, frame rate (i.e., frame per second, FPS) and quantization parameter (QP), respectively.
The objective of configuration-adaptive video streaming is to select the best configuration for each chunk $i\in\{1, 2, ..., N\}$ to maximize the task accuracy $\mathcal{A}(\boldsymbol{c}_i)$ and minimize the upload lag $\mathcal{L}(\boldsymbol{c}_i)$~\cite{zhang2022casva}. The upload lag of a chunk is defined as the difference between its expected upload time and actual upload time. For example, assume that chunk $i$ is uploaded to server at time $t_i$, while its expected upload time\footnote{In this paper, we assume that the start time of a live video streaming is 0.} is $i \times T$, then the lag experienced by chunk $i$ is $t_i - i \times T$. Note that lag is non-negative, i.e., $\mathcal{L}(\boldsymbol{c}_i) = \max(t_i-i \times T,0)$.}

\Duo{Despite its simplicity, the configuration adaption problem encounters two practical challenges. First, there exists a contradiction between the two goals of live VA. Intuitively, selecting an expensive configuration will result in a high accuracy but consume much bandwidth. When the network bandwidth is dynamic and scarce, this may quickly deplete the bandwidth resources and produce great lag, decreasing the efficiency of live VA. 
On the other hand, a cheap configuration may reduce the lag and transmission cost, but the accuracy may suffer due to the degraded video quality.
Second, the lag of chunk $i$ is not only affected by its upload delay $u_i$, but also the lag of its previous chunk $i-1$, i.e., $\mathcal{L}(\boldsymbol{c}_i)=\max(\mathcal{L}(\boldsymbol{c}_{i-1}) + u_i - T, 0)$. Hence, the configuration of a chunk has cascading effects on the subsequent chunks, which suggests the problem is essentially a sequential decision problem. For instance, streaming the current chunk with an unnecessary expensive configuration tends to increase lag, and therefore, the following chunks may be streamed with cheap configurations to achieve live analytics in the risk of accuracy drop. }

\subsection{Methodology}
\Duo{Due to the aforementioned challenges, it is impractical to solve the configuration selection problem with analytical models, which motivates us to design a learning-based method. In particular, we adopt the advanced imitation learning (IL)~\cite{hussein2017imitation} framework to solve the problem, which consists of an agent (represented as a neural network) to select configurations, and an expert model to ``teach'' the agent in the offline training phase. The details are explained as follows.}

\subsubsection{Neural Network Architecture}
\Duo{\textbf{Inputs: }
At each time stamp $t$, the agent will receive a set of observable variables $s_t$, representing the state information of network environment, camera buffer and video content etc. Specifically, we define $s_t$ as follows:
$$ s_t = \{\vec{v_t}, \vec{n_t}, \vec{u_t}, b_t, \vec{r_t},\vec{f_t}, \vec{q_t},M_t \} $$
Here, $\vec{v}_t$ is the vector of past $k$ chunk volumes. $\vec{n}$ is the throughput measurements for past $k$ chunks, which represents the network conditions. $\vec{u_t}$ is the upload delay of past $k$ chunks. $b_t$ represents the camera buffer size, i.e., how many frames (e.g., 2 seconds) are buffered at the camera when uploading chunk $t$. The larger buffer size $b_t$,  the larger accumulative upload lag. $\vec{r}_t, \vec{f}_t, \vec{q}_t$ represent the resolution, FPS and QP selected for past $k$ chunks. Finally, $M_t$ is the motion feature map for chunk $t$, which allows the agent to perceive the content dynamics of videos.}

\noindent \Duo{\textbf{Outputs: }The agent takes the state information $s_t$ as input, and outputs the probability distribution of each configuration. The configuration with the maximum probability will be selected as the action $a_t$ corresponding to $s_t$, which will be sent back to the camera to encode the next video chunk. }

\begin{figure}[t]
    \centering
    \subfigure[Neural network architecture of the agent]{
    \includegraphics[width=0.48\textwidth]{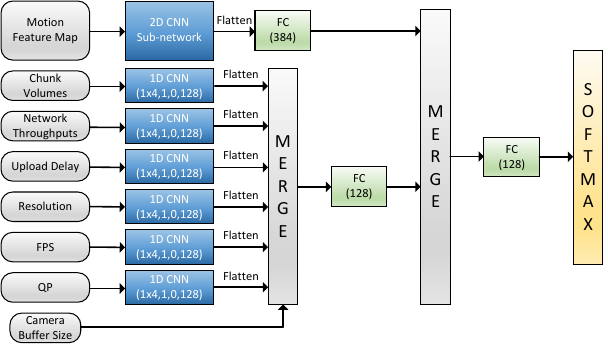}
    \label{subfig:dnn}
    }
    \subfigure[Architecture of 2D CNN sub-network]{
    \includegraphics[width=0.48\textwidth]{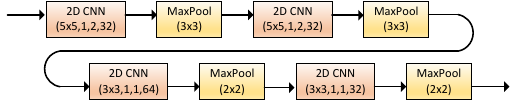}
    \label{subfig:cnn_subnet}
    }
    \caption{Neural network architecture of \texttt{ILCAS} agent. Model parameters: 1D/2D CNN (kernel size, stride, padding, channels), FC (number of neurons), MaxPool (kernel size).}
    \label{fig:dnn_architecture}
\end{figure}

\noindent \Duo{\textbf{Network Architecture: }Figure~\ref{fig:dnn_architecture} depicts the neural network architecture of the agent. For numerical inputs, we first use 1D CNN layers to extract hidden features from each input vector. These features are then flattened, concatenated as a new vector, and fed into a fully connected (FC) layer. Considering that camera buffer size contains more important information for making decisions, especially for controlling lag, we provide a shortcut for input $b_t$ by directly concatenating it with the vector merged by the flattened 1D CNN features. For the input motion feature map, we leverage 2D CNN sub-network (see Figure~\ref{subfig:cnn_subnet}) to extract underlying features of content dynamics, which are next fed into a FC layer. All processing results are merged together, and fed to another FC layer to learn the complex relationship between different features. Finally, a softmax layer is used to output the probability distribution of each configuration.}

\subsubsection{Learning Framework}
\Duo{We adopt the state-of-the-art Generative Adversarial Imitation Learning (GAIL)~\cite{generative2016ho} as the IL framework to train our agent since it outperforms traditional IL frameworks in efficiency and scalability.}

\begin{figure}[t]
  \centering
  \centerline{\epsfig{figure=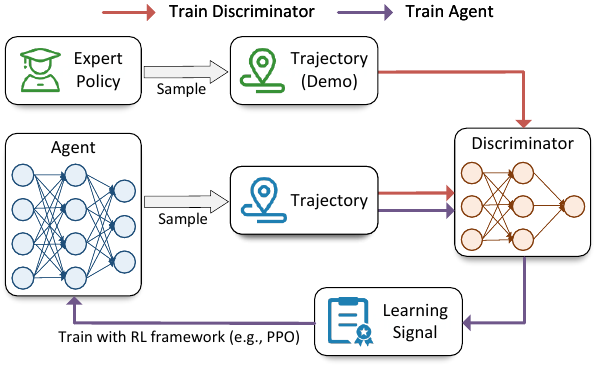,width=0.48\textwidth}}
  \caption{General training procedure of GAIL.}
  \label{fig:gail}
\end{figure}

\Revision{Figure~\ref{fig:gail} illustrates the general training procedure of GAIL. As shown, GAIL consists of three core components: 
\begin{itemize}
    \item \textit{Expert} provides demonstrations of solving the adaptive configuration selection problem for training.
    \item \textit{Discriminator} is used to facilitate the agent to learn the behavior pattern of the expert. It in essence operates as a binary classifier that distinguishes between the expert's and agent's behaviors.
    \item \textit{Agent} learns to imitate the expert's behavior of taking actions (i.e., selecting configurations) with the assistance of the discriminator.
\end{itemize}
The training procedure of GAIL is implemented in the adversarial training manner~\cite{generative2016ho}. Let $\pi_E$ represent the expert policy, and $\pi_\theta$ represent the agent parameterized by $\theta$. Define $D_\omega$ as the discriminator model parameterized by $\omega$. At each training step $t$, GAIL samples a demonstration and trajectory composed of state-action pairs under the expert and agent policy, respectively: $\tau_E=\{\hat{s}_i,\hat{a}_i\}_{i=0}^n \sim \pi_E$, $\tau_t=\{s_i,a_i\}_{i=0}^m \sim \pi_{\theta_t}$. It then uses $\tau_E$ and $\tau_t$ as the training data to update the discriminator $D_{\omega_t}\rightarrow D_{\omega_{t+1}}$ with the binary cross entropy loss function. Next, the discriminator $D_{\omega_{t+1}}$ is used to update the agent $\pi_{\theta_t}\rightarrow\pi_{\theta_{t+1}}$, which guides the agent to learn the behavior pattern of the expert. Specifically, for each state-action pair $(s,a)$ sampled by the agent, the discriminator takes $(s,a)$ as input and computes $D_{\omega_{t+1}}(s,a)$ indicating the probability of the expert taking action $a$ when it encounters state $s$. The higher probability $D_{\omega_{t+1}}(s,a)$ means the agent imitating the expert's behavior better. Hence, the learning signal for training the agent is defined as:
\begin{equation}
p(s,a)=\log D_{\omega_{t+1}}(s,a)
\end{equation}
$p(s,a)$ can be regarded as the reward signal in RL. Thus, based on the trajectory of $(s, a, p(s, a))$, the agent can be updated using common RL framework such as Proximal Policy Optimization (PPO)~\cite{schulman2017proximal}. Note that both the expert and discriminator are designed to guide the training of agent, and thus they are used only in the offline training phase.}

\begin{algorithm}[t]
\caption{Expert}
\label{algo:expert}
    \KwIn {$N$: chunk number; $\mathcal{C}$: set of all possible configurations; $L$: maximum acceptable lag}
    \KwOut {$\mathbb{C}$: configurations for each chunk}
    \tcp{Initialization}
    Initialize matrix $A$ with $0$, matrix $S$ and $Q$ with $\emptyset$. \\
    \ForEach{$\boldsymbol{c} \in \mathcal{C}$}{
        $l = \max(0, \mathcal{U}_1(\boldsymbol{c}) - T)$ \\
        \If{$l \le L$ and $\mathcal{A}_1(\boldsymbol{c}) > A[1, l]$} {
            $A[1, l] \leftarrow \mathcal{A}_1(\boldsymbol{c})$ \\
            $S[1, l] \leftarrow \boldsymbol{c}$
        }
    }
    \tcp{Update}
    \For{$i\leftarrow 1$ to $N-1$} {
        \For{$j\leftarrow L$ to $0$} {
            \If{$S[i, j] = \emptyset$}{
                continue. \tcp{Skip invalid entries}
            }
            \ForEach{$\boldsymbol{c} \in \mathcal{C}$} { 
                $l = \max(0, j + \mathcal{U}_{i+1}(\boldsymbol{c}) - T)$ \\
                \If{$l \le L$ and $A[i, j] + \mathcal{A}_{i+1}(\boldsymbol{c}) > A[i + 1, l]$} {
                    $A[i + 1, l] \leftarrow A[i,j] + \mathcal{A}_{i+1}(\boldsymbol{c})$\\
                    $S[i + 1, l] \leftarrow \boldsymbol{c}$ \\
                    $Q[i + 1, l] \leftarrow (i , j)$
                }
            }
        }
    } 
    $\mathbb{C} \leftarrow \{\}$\\
    $i \leftarrow N$\\ $j \leftarrow \mathop{\arg\max}\limits_{l \in [0, L]} A(N, l)$ \\
    \While{$i > 0$} {
        $\mathbb{C} \cup (i, S[i, j])$ \\
        $i, j \leftarrow Q[i, j]$
    }
    \Return{$\mathbb{C}$}
\end{algorithm}

\subsubsection{Expert Design}
\Duo{The key of implementing GAIL in our problem is the design of discriminator and expert. Following the practice in~\cite{generative2016ho}, the discriminator can share the same architecture with the agent with minor modifications: the second merged features in Figure~\ref{subfig:dnn} are concatenated with the action one-hot vector, and the output layer is replaced by sigmoid layer. Therefore, the technical challenge here is the expert design.}

\Duo{Recall that the optimization problem of adaptive configuration selection for live VA is to select the best configuration for each chunk to maximize VA accuracy and minimize the upload lag. This problem has no analytical solutions due to the intrinsic conflicts between two goals. 
Fortunately, keeping the upload lag under a certain threshold is acceptable in practice. The rationale behind is that different applications have different lag goals, meaning they have tolerance on lag to some degrees~\cite{zhang2017live}. For instance, lagging tens of seconds to upload the video is even acceptable for license plate reader at toll route~\cite{sr520bridgetolling} because the billing can be delayed.
Based on this observation, the aforementioned problem can be relaxed as finding the configuration for each chunk to maximize the accuracy while keeping the upload lag under $L$. Here, $L$ represents the maximum acceptable upload lag, which can be easily determined according to specific applications. 
}

\Duo{
As shown in Algorithm~\ref{algo:expert}, we design the expert with an efficient dynamic programming algorithm to solve the problem.
Matrix $A[i, j]$ records the maximum accuracy of chunk $i$ when upload lag is $j$, $S[i, j]$ records the configuration when $A[i,j]$ achieves the maximum, and $Q[i,j]$ records the previous entry that transits to $S[i,j]$.
Variable $T$ is the length of each chunk. Since the expert is an offline policy, we assume that it has the full knowledge of configuration performance and network conditions, and thus is aware of the accuracy $\mathcal{A}_i(\boldsymbol{c})$ and upload delay $\mathcal{U}_i(\boldsymbol{c})$ for chunk $i$ with configuration $\boldsymbol{c}$.
The main idea of the algorithm is to iterate over all possible lag for each chunk, and select the configuration with the maximum accuracy under each situation. Line 1-16 compute matrix $S$ and matrix $Q$, then line 17-22 search for the optimal configuration for each chunk according to $S$ and $Q$ through backtracking.
The output of the algorithm is a set of selected configurations $\mathbb{C}$ (line 23).
The expert's demonstration composed of state-action pairs can be obtained by replaying the video with the selected configuration set $\mathbb{C}$. The detailed process is omitted here for simplicity.}

\Revision{Note that the expert is infeasible to online adjust configurations for live VA streaming, as it requires the streaming of chunks in golden configuration to profile the accuracy of each configuration as well as the perfect knowledge of network conditions.} The role of expert is to guide the training of agent, and therefore it is used only in the offline training phase. Besides, the upload lag is originally a continuous variable, but in our practical implementation of Algorithm~\ref{algo:expert}, we discretize this variable with certain granularity (e.g., using every 0.1 second as a gap), since two very close lags will not make a big difference.

\Revision{The time complexity of expert algorithm is $O(NL|\mathcal{C}|)$, where $N$ is the chunk number, $L$ is the maximum acceptable upload lag, and $|\mathcal{C}|$ is the number of all possible configurations. In the later experiment settings (see Section~\ref{subsec:evaluation_setup}), we find that the runtime of expert to sample one demonstration is less than 1 second. Therefore, the running cost of expert introduced to the training process is negligible.}

\section{Cross-camera Collaboration}
\label{sec:cross_camera_collaboration}
In this section, we introduce the detailed design of the cross-camera collaboration scheme in \texttt{ILCAS}. We first quantify the spatio-temporal correlations across cameras and then elaborate how to utilize the camera correlations to achieve more accurate configuration adaption.
\subsection{Spatio-Temporal Correlations}

\noindent\textbf{Spatial correlations:} The trajectories of objects in camera networks show a strong correlation, which indicates a high degree of spatial correlation of cameras. The degree of spatial correlation $\mathbb{S}$ between source camera $c_s$ and target camera $c_d$ can be quantified by the ratio of: (a) the number of objects appearing in both the source camera's FOV and the target camera's FOV $O_{c_s,c_d}$, to (b) the number of objects appearing in the source camera's FOV $O_{c_s}$~\cite{jain2020spatula}:

\begin{equation}
\mathbb{S}(c_{s}, c_{d}) = \frac{\sum{O_{c_{s},c_{d}}}}{\sum{O_{c_{s}}}}
\end{equation}

\noindent\textbf{Temporal correlations:} 
The traveling time of objects between camera pairs reveals their temporal correlations over time. If a large portion of objects leaving source camera $c_s$ to target camera $c_d$ arrives within a certain time duration $[t_1, t_2]$, the camera $c_d$ is highly correlated to the camera $c_s$ in the time window $[t_1, t_2]$. The degree of temporal correlations $\mathbb{T}$ can be quantified by the ratio of: (a) objects arriving $c_d$ from $c_s$ in a certain time window $[t_1, t_2]$ to (b) total objects arriving $c_d$ from $c_s$~\cite{jain2020spatula}:

\begin{equation}
\mathbb{T}(c_{s}, c_{d}, [t_1, t_2]) = \frac{\sum_{t=t_1}^{t_2}{O_{c_{s},c_{d}}}}{\sum{O_{c_{s},c_{d}}}}
\end{equation}

\begin{figure}[t]
    \centering
    \begin{minipage}[t]{0.5\textwidth}
    \subfigure[Motion feature map w/o filter]{
    \includegraphics[width=0.45\textwidth]{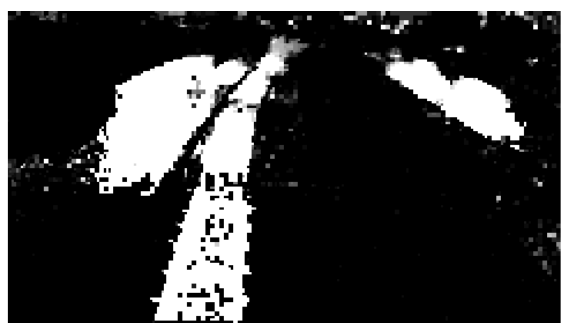}
    \label{subfig:cross_mf_wo_filter}
    }
    \subfigure[Motion feature map with filter]{
    \includegraphics[width=0.45\textwidth]{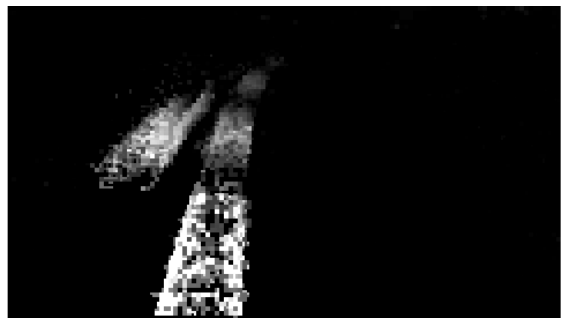}
    \label{subfig:cross_mf_with_filter}
    }
    \end{minipage}
    \subfigure[Directions of camera's FOV]{
    \includegraphics[width=0.3\textwidth]{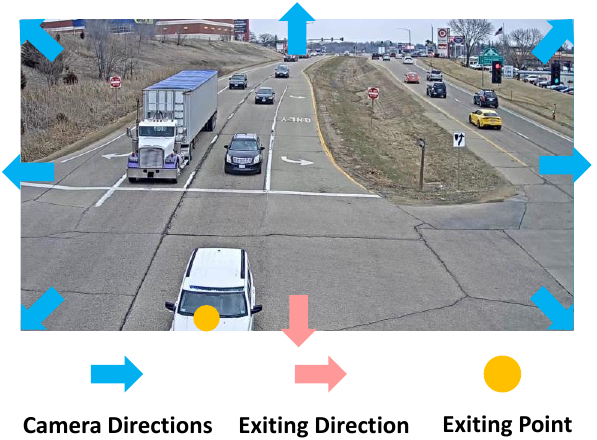}
    \label{subfig:cross_camera_directions}
    }
    \caption{Illustration of directions and motion feature map filter.}
    \label{fig:mf_filter}
\end{figure}

\subsection{Cross-Camera Collaboration Design}

\textbf{Info-sharing mechanism:} As described in Section~\ref{sec:cross_camera_enhancement}, the key to cross-camera-based configuration adaption is to utilize the information from correlated cameras. To this end, we design an info-sharing mechanism to share the motion feature maps across cameras. 
\Revision{As a target camera may have many source cameras, some of which may not be highly-correlated, we set a spatial threshold $\mathbb{S}_{thresh}$ and a temporal threshold $\mathbb{T}_{thresh}$ to filter out low-correlated cameras.}
\Revision{Given a target camera $c_d$, we search for the source camera $c_s$ within the time window $[t_1, t_2]$, which satisfies $\mathbb{S}(c_{s}, c_{d}) \geq \mathbb{S}_{thresh}$ and $\mathbb{T}(c_{s}, c_{d}, [t_1, t_2]) \geq \mathbb{T}_{thresh}$. 
Cameras that do not meet these conditions are regarded as low-correlated cameras, which are filtered out and prevented from sharing information with the target camera. Finally,}
for the source camera $c_s$ which meets the aforementioned conditions, $c_d$ can utilize the motion feature maps from $c_s$ in the certain time duration $[t_1, t_2]$. Specifically, as camera video is organized as video chunks to transmit, we collect the motion feature maps from $c_s$ at the past chunk $p$ where $p\in[t_1, t_2]$.

\noindent\textbf{Motion feature map filter:} 
Motion feature map is a piece of important information for \texttt{ILCAS}'s agent to select configurations as it reflects video content dynamics, but directly applying the raw feature maps from source cameras to target camera is infeasible. This is because the noise in raw feature maps from source cameras (e.g., redundant information about objects that will not move to the target camera) may misleads the agent to select inappropriate configurations.
Therefore, we design a motion feature map filter for noise filtering. As shown in Figure~\ref{fig:mf_filter}, we divide the camera's FOV in 8 directions every $45^{\circ}$. We denote $D_{c_{s},c_{d}}$ as the exiting direction of objects coming from source camera to target camera, $(X_D, Y_D)$ as the exiting point on the motion feature map and $D(V_{x, y})$ as the direction of the motion degree $V$ at pixel $x, y$. To filter out the noisy information, we only save motion degrees that have the same direction as $D_{c_{s},c_{d}}$. 
Besides, we also apply a linear decay method on the motion feature map according to the distance between each pixel and the existing point. The rationale behind is that objects located nearer the existing point will appear in the target camera's FOV sooner, and vice versa.
To summarize, for the motion feature map with $X$ rows and $Y$ columns, the design of filter is as follows:

\begin{small}
\begin{gather}
	V_{x, y}=\left\{
	\begin{aligned}
		&V_{x, y}\! \times\! (1\! -\! \frac{\sqrt{(X_D\! -\! x)^2\! + (Y_D\! -\! y)^2}}{\sqrt{(X\! -\! X_D)^2\! +\! (Y\! -\! Y_D)^2}}),\! D(V_{x, y}) \!=\! D_{c_{s},c_{d}} \\
		&0, \quad  D(V_{x, y}) \neq D_{c_{s},c_{d}} 
	\end{aligned}
	\right.
\end{gather}
\end{small}

\noindent\textbf{Cross-camera feature extraction model:}
After filtering, the motion feature maps collected from source camera now can be used to infer future video dynamics for the target camera. To extract hidden features from these feature maps, we introduce Convolutional Long Short-Term Memory (ConvLSTM)~\cite{ConvLSTM} as the feature extraction model, which has a strong ability in extracting spatio-temporal correlations and is widely used in forecasting problems. We collect past motion feature maps from source camera and target camera, then organize them as feature map groups, respectively. 
\Revision{The ConvLSTM then extracts the spatio-temporal features from the feature map groups and outputs the refined feature maps. These refined feature maps are essentially the motion feature maps predicted by ConvLSTM, which reflects the future video dynamics more accurately. Next, the outputs of ConvLSTM are fed into the agent (i.e., the 2D CNN sub-network to process motion feature maps, as shown in Figure~\ref{fig:dnn_architecture}) for more proper configuration selection.}

\section{Evaluation}
\label{sec:Evaluation}

\begin{table*}[t]
    \caption{Video dataset information.}
    \label{tab:video}
    \vspace{-0.2cm}
    \centering
    \begin{tabular}{m{2cm}<{\centering} m{4cm}<{\centering} m{8.5cm}<{\centering}}
        \toprule
         \textbf{Video} & \textbf{Type} & \textbf{Description} \\
         \midrule
         \textit{Dash1}~\cite{dash1} & Dash camera & Driving downtown in Chicago at daytime.\\
         \midrule
         \textit{Dash2}~\cite{dash2} & Dash camera & Driving downtown in London at night.\\
         \midrule
         \textit{Stationary1}~\cite{stationary1} & Stationary traffic camera & Road traffic video recorded on a sunny day.\\
         \midrule
         \textit{Stationary2}~\cite{stationary2} & Stationary traffic camera & Highway traffic video recorded on a sunny day. \\
         \midrule
         \textit{Stationary3}~\cite{stationary3} & Stationary traffic camera & Road traffic video recorded at sunset.\\
         \midrule
         \textit{AI City}~\cite{cross} & Stationary traffic camera & Video collected from multiple cameras located at an intersection.\\
         \bottomrule
    \end{tabular}
\end{table*}

\subsection{Evaluation Setup}
\label{subsec:evaluation_setup}
\textbf{Vision task and video datasets:} We take object detection, a typical video analytic task, as an example to evaluate \texttt{ILCAS}. We use the pretrained FasterRCNN-ResNet101~\cite{ren2015faster} as the detection model. We collect five video clips as the evaluation dataset from YouTube, and use a cross camera video dataset (collected from 2022 AI City Challenge~\cite{cross}) to evaluate the effectiveness of the proposed cross-camera collaboration scheme. The detailed information of video datasets is presented in Table~\ref{tab:video}. Unless otherwise noted, the chunk length of each video is set to 1 second, i.e. $T=1$. For each video, we use the first $80\%$ of chunks for training while the remaining $20\%$ for testing.

\noindent\textbf{Network traces:} \Revision{Since there are no existing bandwidth datasets collected from the camera side, we use a public 4G/LTE bandwidth dataset~\cite{van2016http} as an alternative to simulate real-world network conditions for evaluation. However, the mean bandwidth throughputs of all traces in the dataset are too high (about 30Mbps)~\cite{van2016http}, while in practice the bandwidth resources between the cameras and servers are limited (e.g., 1Mbps in average)~\cite{zhang2018awstream}. To this end, we scale the traces down to a lower range of (0.2, 2.0) Mbps to match the practical real-world network conditions of cameras. To utilize the fluctuation patterns of the bandwidth traces, we scale the traces while maintaining their variances. This method of preprocessing the bandwidth dataset is also adopted in~\cite{zhang2022casva}.}
\Revision{When conducting cross-camera evaluation, we treat the network environment of each camera as an independent system that does not affect the others. This is because there is no direct information transmission between cameras since motion feature maps are uploaded to servers instead of direct transmission across cameras, as shown in Figure~\ref{fig:system}.}

\noindent\textbf{IL training settings:} We empirically feed past $k=8$ sample information into our agent. We use PPO~\cite{schulman2017proximal} framework to update agent's parameters during IL training. By default, the learning rate is $0.0001$, the discount factor is $0.95$, the clipped parameter is $0.2$ and the entropy coefficient is $0.02$. We consider a configuration as the combination of three knobs: resolution $r\in \{1, 0.8, 0.6, 0.5, 0.4, 0.3\}$\footnote{By multiplying $r$ with the original resolution of a video, we will obtain the resolution of the resized video. For example, $1080\text{p} \times 0.5 = 540\text{p}$.}, FPS $f\in \{30, 15, 10, 5, 2, 1\}$ and QP $q\in \{21, 25, 29, 33, 37, 41\}$. This results in a set of 216 distinct configurations and thus the agent's output dimension is 216. The max acceptable upload lag of expert is empirically set to 1 second, i.e., $L=1$.

We use PyTorch\footnote{https://pytorch.org/} and Tianshou\footnote{https://tianshou.readthedocs.io/en/master/} to implement DNN and PPO, respectively. All experiments are performed on a trace-driven simulator, which streams video chunks from a camera client to a server. The RTT time between the camera and server is set to 80ms. \Revision{We run\footnote{\Revision{The computing resources we use to run our experiments are highly redundant.}} DNN inference on RTX 3080Ti and other computations on Intel Xeon Silver 4210.}

\noindent \textbf{Evaluation metrics:} We consider \textit{mean accuracy}, \textit{mean lag}, \textit{accuracy CDF} and \textit{lag CDF} as evaluation metrics. The mean accuracy or lag is calculated as the mean inference accuracy or upload lag of all test video chunks.

\noindent\textbf{Baselines:} We compare \texttt{ILCAS} with the following baselines, which stand for the state-of-the-art profiling-based and DRL-based solutions:
\begin{itemize}
    \item \textit{Chameleon}~\cite{jiang2018chameleon}: This approach divides a streaming session into sequential windows, and each window is further divided into several segments. 
    It periodically profiles a set of candidate configurations in the first segment of each window, and derives the best configurations from the profiled ones for the remaining segments in the window.
    \item \textit{CASVA}~\cite{zhang2022casva}: This approach employs DRL to train an agent for configuration selection, with the objective of maximizing task accuracy while minimizing upload lag. It uses a handcrafted reward function during the training process and also takes into account the volume changes of the current and previous chunks in the same configuration as an indicator of content dynamics.
    \item \textit{Expert}: We sample a demonstration under the expert policy described in Section~\ref{sec:il_based_configuration_selection} to serve as the offline optimal result.
\end{itemize}

\begin{figure*}[t]
    \centering
    \subfigure[Mean accuracy]{
    \includegraphics[width=0.46\textwidth]{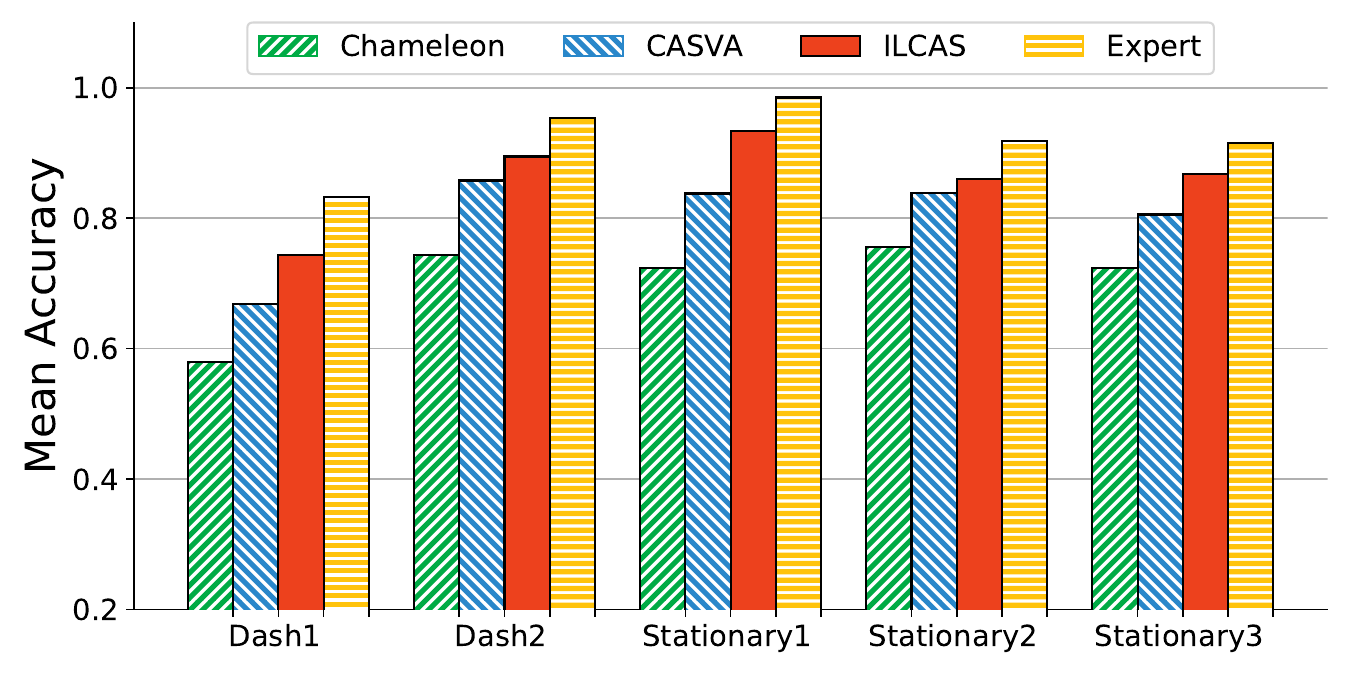}
    \label{subfig:mean_acc}
    \vspace{-0.5cm}
    }
    \hspace{0.5cm}
    \subfigure[Mean lag]{
    \includegraphics[width=0.46\textwidth]{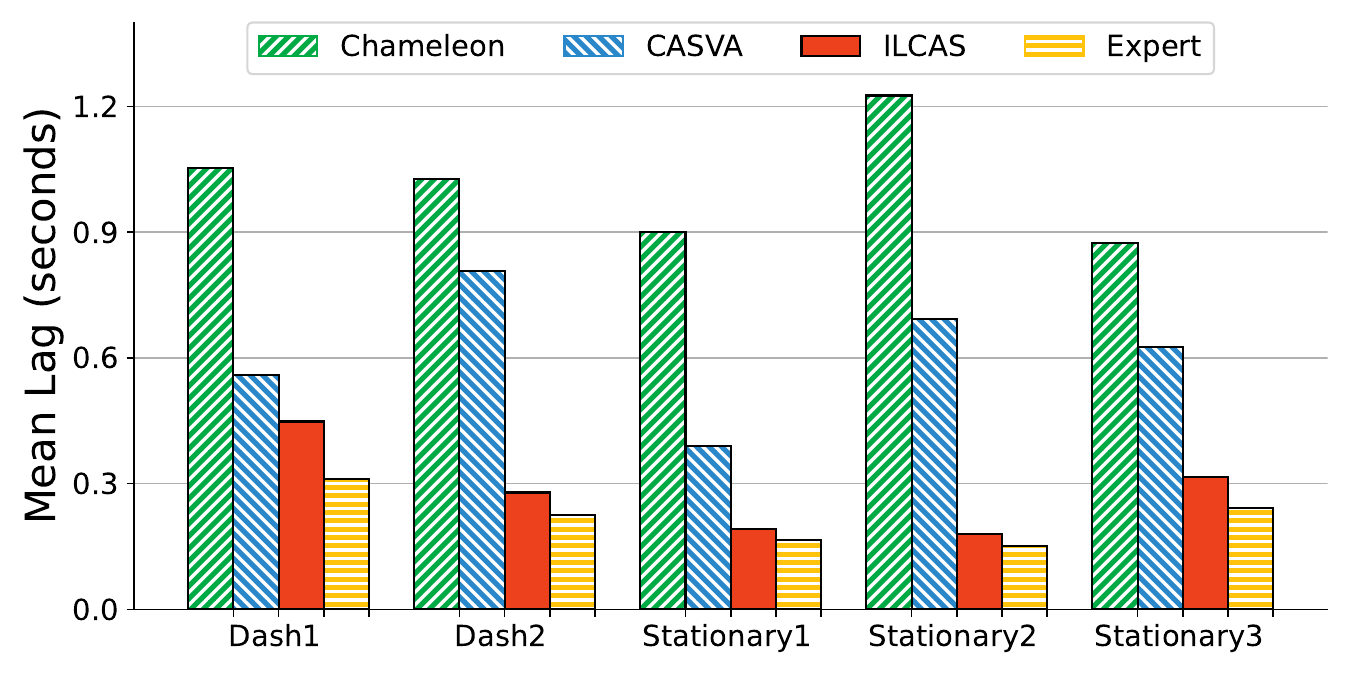}
    \label{subfig:mean_lag_gap}
    \vspace{-0.5cm}
    }
    \vspace{-0.3cm}
    \caption{The performance of different methods on various videos.}
    \label{fig:mean_acc_lag_methods}
\end{figure*}

\begin{figure}[t]
    \centering
    \subfigure[Accuracy CDF]{
    \begin{minipage}[b]{0.165\textheight}
    \includegraphics[width=\textwidth]{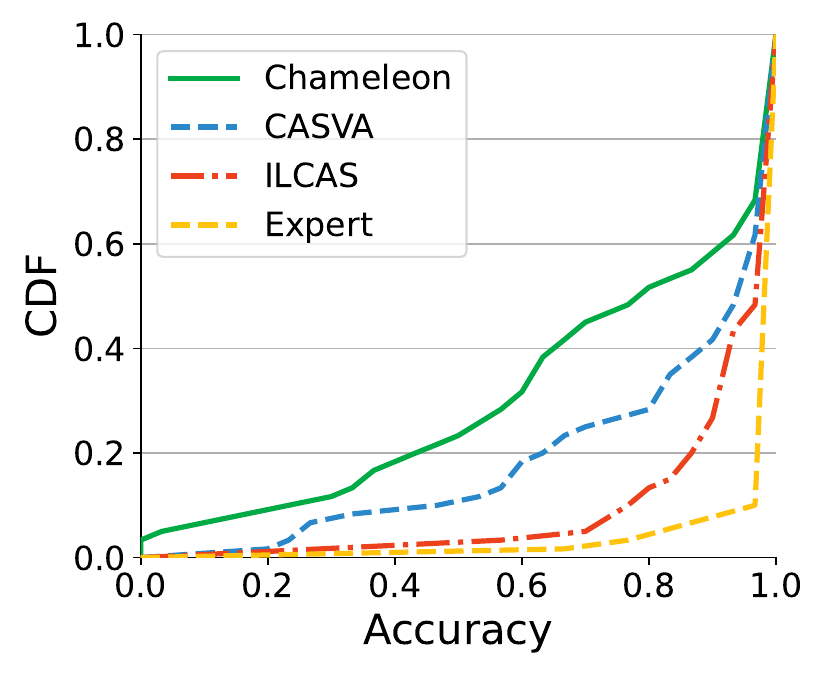}
    \label{subfig:acc_cdf}
    \vspace{-0.5cm}
    \end{minipage}
    }
    \subfigure[Lag CDF]{
    \begin{minipage}[b]{0.165\textheight}
    \includegraphics[width=\textwidth]{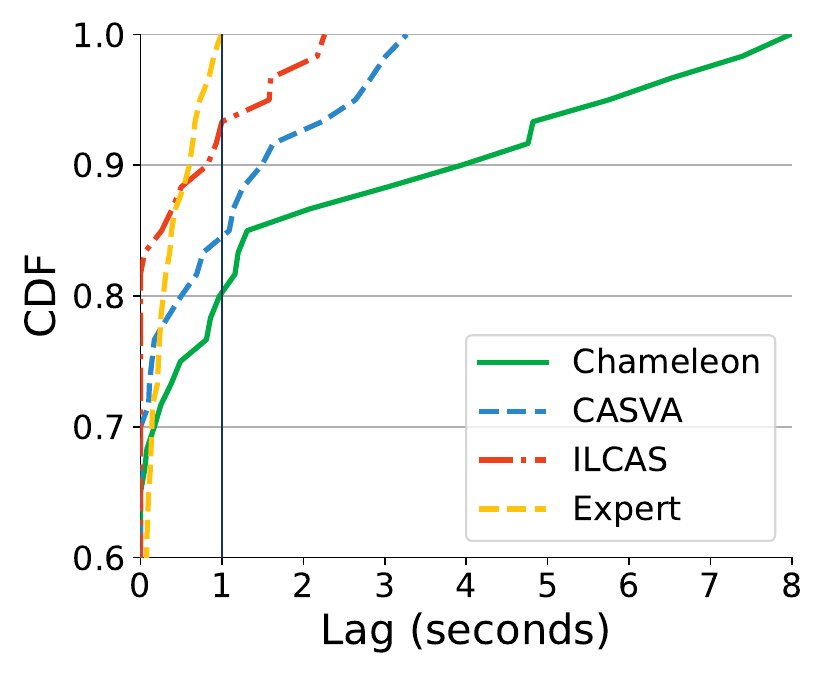}
    \label{subfig:lag_cdf}
    \vspace{-0.5cm}
    \end{minipage}
    }
    \vspace{-0.3cm}
    \caption{The performance of different methods on \textit{Stationary1} video.}
    \label{fig:enter-label}
\end{figure}

\subsection{\texttt{ILCAS} vs. Baselines}
\label{subsec:ILCAS_vs_baselines}
Figure~\ref{fig:mean_acc_lag_methods} compares the performance of different methods on various videos. It can be seen that \textit{Chameleon} achieves the lowest mean accuracy while producing the highest mean lag. This is because the candidate configurations profiled by \textit{Chameleon} may not work well for the whole time window. When the video contents or network conditions suddenly change, the profiled configurations can produce low accuracy or high lag, but \textit{Chameleon} has to trigger a new profiling process to update candidate configurations, which only achieves coarse-grained adaption. 

Compared to \textit{Chameleon}, \textit{CASVA} achieves higher accuracy and lower lag since it adopts an DRL-based learning strategy for selecting configurations to better adapt to environment changes.  \Revision{However, there still exists a significant performance gap between \textit{CASVA} and \textit{Expert}. In contrast, \texttt{ILCAS} exhibits the smallest gap with the \textit{Expert}, indicating its superiority over \textit{CASVA}. The performance gain of \texttt{ILCAS} over \textit{CASVA} can be attributed to two key aspects.}
\begin{itemize}
    \item \Revision{Firstly, \textit{CASVA} relies on chunk volume changes as the indicator to infer content changes. However, this indicator only provides coarse-grained information and contains much noise, which may mislead \textit{CASVA} to select inappropriate configurations. On the other hand, \texttt{ILCAS} leverages motion feature maps to effectively capture the fine-grained changes of video contents. This enables \texttt{ILCAS} to efficiently adapt to video dynamics and make more appropriate configuration selections.}
    \item \Revision{Secondly, \textit{CASVA} employs a fixed reward function for training, which fails to properly craft the trade-off between accuracy and upload lag across different scenarios. In contrast, \texttt{ILCAS} leverages the advanced IL framework to train the agent. It efficiently utilizes the expert as the offline optimal policy to guide the agent in selecting configurations according to environment changes such as video contents and network conditions, thereby striking a good balance between VA accuracy and upload lag.}
\end{itemize}

\Revision{As a result, \texttt{ILCAS} achieves the best trade-off between accuracy and upload lag compared to \textit{Chameleon} and \textit{CASVA}. As depicted in Figure 11, \texttt{ILCAS} improves the mean accuracy by $15.0\text{--}20.9\%$ and $2.0\text{--}9.6\%$ compared to \textit{Chameleon} and \textit{CASVA}, respectively. Meanwhile, it significantly reduces the mean lag by $57.4\text{--}85.3\%$ and $19.9\text{--}74.0\%$.} Figure~\ref{subfig:acc_cdf} and Figure~\ref{subfig:lag_cdf} further compares the CDF of accuracy and lag of different methods over \textit{Stationary1} video. \Revision{As shown, a large proportion of \texttt{ILCAS} is concentrated in the range of higher accuracy and lower lag, demonstrating the superiority of \texttt{ILCAS} over \textit{Chameleon} and \textit{CASVA}.} Specifically, as illustrated in Figure~\ref{subfig:acc_cdf}, there are $73.3\%$ of chunks with the accuracy above 0.9 for \texttt{ILCAS}, while this value decreases to only $58.3\%$ and $41.7\%$ for \textit{CASVA} and \textit{Chameleon}, respectively. From Figure~\ref{subfig:lag_cdf}, we can see that \texttt{ILCAS} successfully controls the upload lag under 1 second for $93.3\%$ of chunks, much higher than \textit{CASVA} and \textit{Chameleon}. 

\begin{figure}[t]
    \centering
    \subfigure[Mean accuracy]{
    \begin{minipage}[b]{0.165\textheight}
    \includegraphics[width=\textwidth]{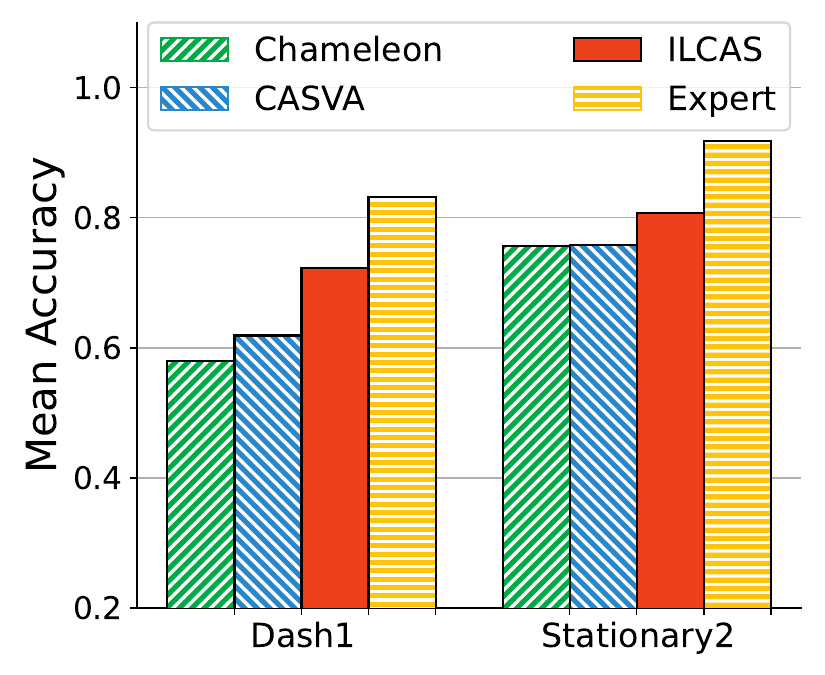}
    \label{subfig:lag_distribute}
    \vspace{-0.5cm}
    \end{minipage}
    }
    \subfigure[Mean lag]{
    \begin{minipage}[b]{0.165\textheight}
    \includegraphics[width=\textwidth]{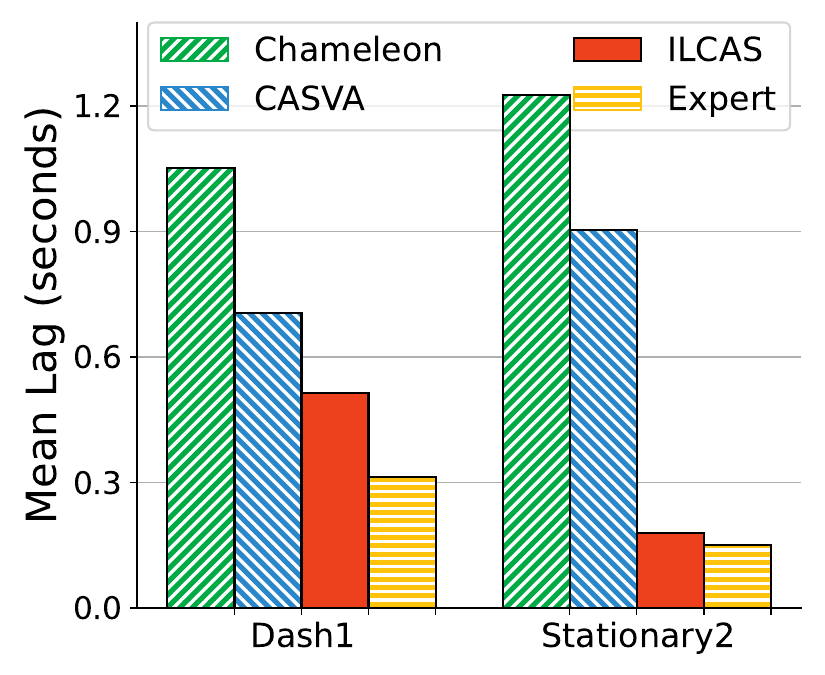}
    \label{subfig:lag_distribute}
    \vspace{-0.5cm}
    \end{minipage}
    }
    \vspace{-0.3cm}
    \caption{\Revision{The generalization performance of learning-based methods \textit{CASVA} and \texttt{ILCAS} when trained on \textit{Stationary1} video and tested on \textit{Dash1} and \textit{Stationary2} videos. The results of non-learning-based methods \textit{Chameleon} and \textit{Expert} on \textit{Dash1} and \textit{Stationary2} videos are also presented for comparative analysis.}}
    \label{fig:generalization}
\end{figure}

\begin{figure}[t]
    \centering
    \subfigure[Mean accuracy]{
    \includegraphics[width=0.4\textwidth]{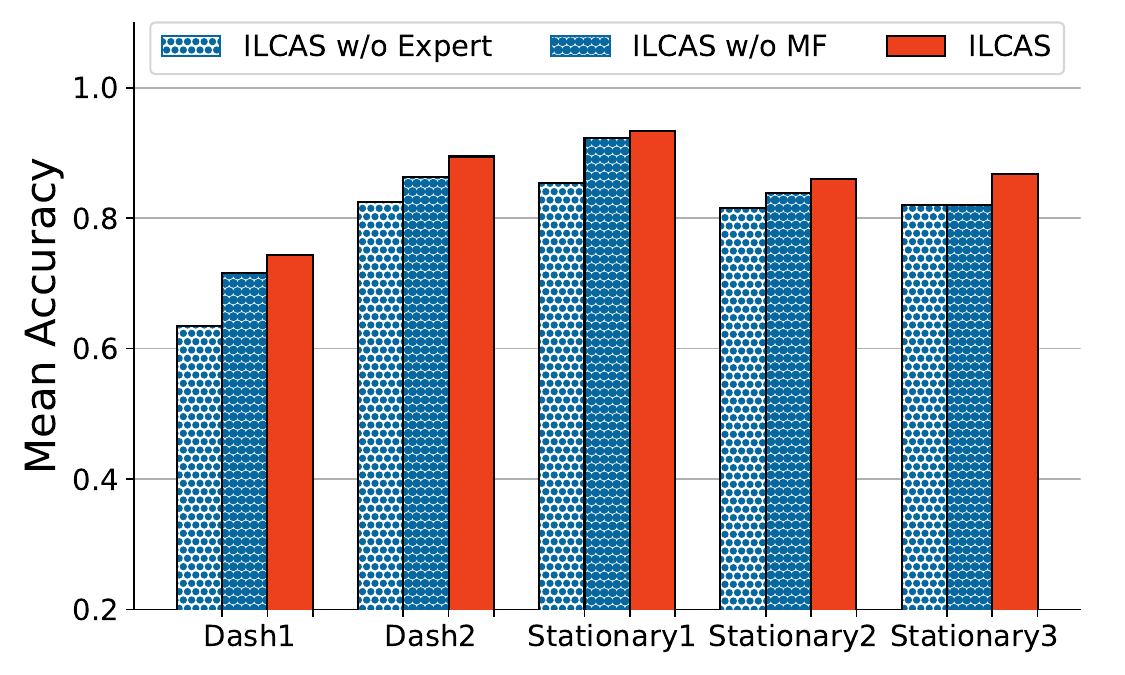}
    \label{subfig:ablation_mean_acc}
    \vspace{-0.5cm}
    }
    \hspace{0.5cm}
    \subfigure[Mean lag]{
    \includegraphics[width=0.4\textwidth]{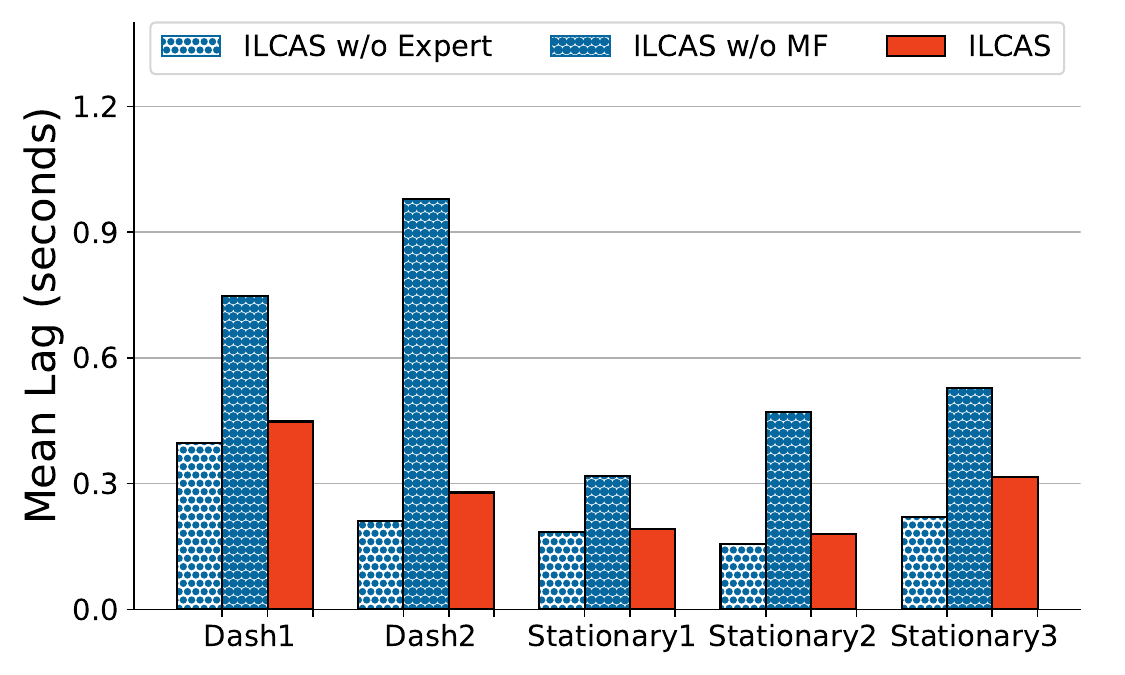}
    \label{subfig:ablation_mean_lag}
    \vspace{-0.5cm}
    }
    \caption{The performance of \texttt{ILCAS} \textit{w/o Expert}, \texttt{ILCAS} \textit{w/o MF} and \texttt{ILCAS} on various videos.}
    \label{fig:ablation_mean}
\end{figure}

\begin{figure}[t]
    \centering
    \includegraphics[width=0.4\textwidth]{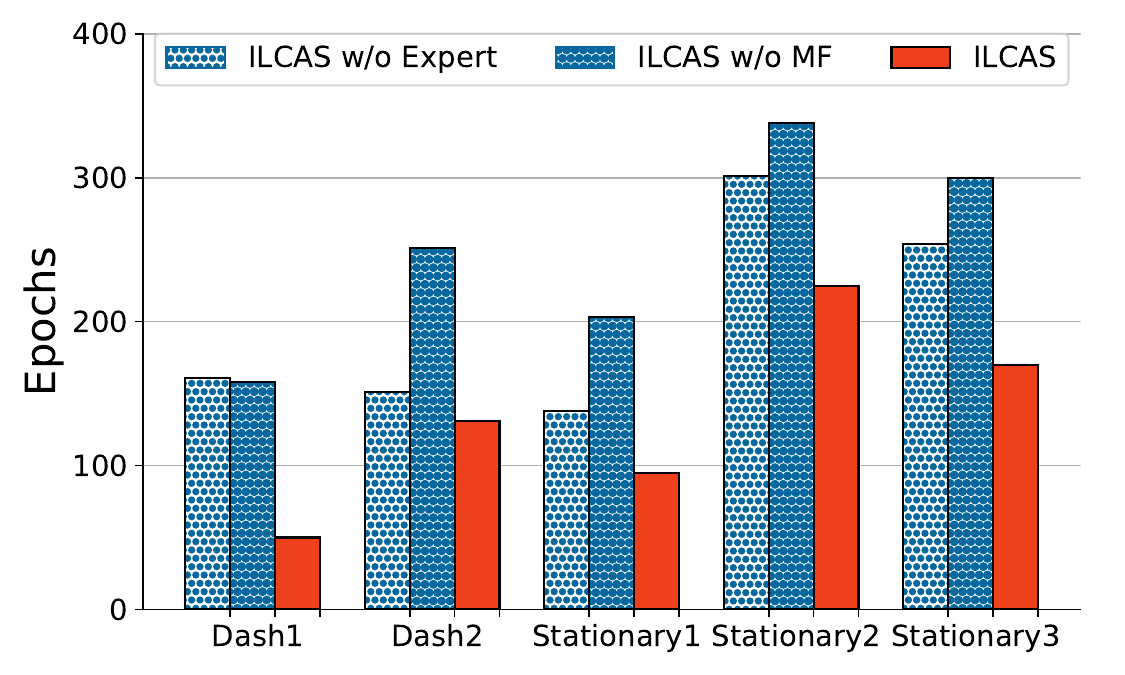}
    \vspace{-0.3cm}
    \caption{\Revision{Number of epochs for \texttt{ILCAS} \textit{w/o Expert}, \texttt{ILCAS} \textit{w/o MF} and \texttt{ILCAS}  to reach certain mean accuracy and lag thresholds.}}
    \label{fig:ablation_epoch_cmp}
\end{figure}

\noindent\Revision{\textbf{Generalization}. In order to evaluate the generalization performance of learning-based methods, we conduct a case study where both \textit{CASVA} and \texttt{ILCAS} are trained on \textit{Stationary1} video and tested on \textit{Dash1} and \textit{Stationary2} videos. The results are reported in Figure~\ref{fig:generalization}. For comparative analysis, we also present the results of \textit{Chameleon} and \textit{Expert} on the testing videos in Figure~\ref{fig:generalization}. Since \textit{Chameleon} and \textit{Expert} are non-learning-based methods which do not require a training process, their results remain the same as those reported in Figure~\ref{fig:mean_acc_lag_methods}}.

\Revision{As depicted in Figure~\ref{fig:generalization}, \texttt{ILCAS} still outperforms \textit{Chameleon} and \textit{CASVA} and achieves the smallest performance gap with the \textit{Expert}. For instance, on \textit{Dash1} video, \texttt{ILCAS} improves the mean accuracy by $14.4\%/10.4\%$ and reduces the mean lag by $51.2\%/27.2\%$ compared to \textit{Chameleon}/\textit{CASVA}. This demonstrates the superior generalization performance of \texttt{ILCAS}. The superiority of \texttt{ILCAS} can be attributed to the common knowledge it learns from the expert, such as selecting cheap configurations under poor network environment and adapting configurations based on video content changes. The learned common knowledge enables \texttt{ILCAS} to generalize across different videos. This indicates that it is possible to use \texttt{ILCAS} to serve multiple cameras with one configuration selection agent instead of training separate ones for each camera.}

\subsection{Ablation Study}
In this part, we set up several experiments to provide a thorough understanding of \texttt{ILCAS}, including the effectiveness of IL framework and motion feature map, as well as the impacts of parameter $L$ on the performance of \texttt{ILCAS}.

\noindent \textbf{Effectiveness of IL.} We remove the IL framework from \texttt{ILCAS} to explore its contributions to the performance of \texttt{ILCAS}. \Revision{Specifically, we train the agent without the guidance of the expert, which forms the method \texttt{ILCAS}\textit{ w/o Expert}. The agent of \texttt{ILCAS}\textit{ w/o Expert} is trained through the reinforcement learning framework with the reward function defined as follows:}
$$reward = \alpha \times Accuracy - (1 - \alpha) \times Lag$$
where $\alpha \in [0, 1]$. By default, $\alpha =0.5$.

Figure~\ref{fig:ablation_mean} depicts the performance of \texttt{ILCAS} and \texttt{ILCAS}\textit{ w/o Expert}. As shown, without expert, the mean lag decreases to only a limited extent but at the cost of accuracy drop. This is because without the expert to demonstrate how to solve the task, \texttt{ILCAS}\textit{ w/o Expert} fails to achieve a good trade-off between task accuracy and upload lag due to the inappropriate definition of reward function. As a result, it tends to conservatively pick up configurations that produce low lag but in sacrifice of task accuracy. For instance, for \textit{Dash1} video, the mean accuracy of \texttt{ILCAS}\textit{ w/o Expert} only reaches $63.4\%$, more than $10\%$ lower than that of \texttt{ILCAS}. On the other hand, it marginally reduces the mean lag by negligibly 0.05s compared to \texttt{ILCAS}. This indicates that enabled by IL, \texttt{ILCAS} selects configurations more gracefully by imitating the expert's behavior and therefore strikes a good balance between task accuracy and upload lag.

\noindent \textbf{Effectiveness of motion feature map.} We next compares the performance of \texttt{ILCAS} and \texttt{ILCAS} \textit{w/o MF} to evaluate the effectiveness of motion feature map. To be more specific, \texttt{ILCAS} \textit{w/o MF} trains the agent with IL, but the motion feature map is excluded from the agent's input.

As illustrated in Figure~\ref{fig:ablation_mean}, \texttt{ILCAS} \textit{w/o MF} generally produces much higher lag and meanwhile achieves lower accuracy than \texttt{ILCAS}. For example, for \textit{Dash2} video, \texttt{ILCAS} \textit{w/o MF} causes about $3\times$ the mean lag of \texttt{ILCAS}, while its mean accuracy is still $3.1\%$ lower than that of \texttt{ILCAS}. The reason behind is that \texttt{ILCAS} \textit{w/o MF} fails to adapt to the dynamics of video contents. On one hand, it chooses cheap configurations for chunks with fast scene changes, which results in low accuracy. On the other hand, it selects unnecessary expensive configurations for chunks with small content changes, producing high lag. By contrast, \texttt{ILCAS} is able to visually ``perceive'' content changes with motion feature map, and thus it succeeds in selecting appropriate configurations that match the video dynamics.

\noindent \Revision{\textbf{Convergence comparison.} As a supplement, we also compare the convergence speed of \texttt{ILCAS} \textit{w/o Expert}, \texttt{ILCAS} \textit{w/o MF} and \texttt{ILCAS}. Figure~\ref{fig:ablation_epoch_cmp} plots the number of epochs for each method to reach a certain accuracy threshold while the mean lag is under 1 second. We set the accuracy threshold as $0.6$ for \textit{Dash1} video and $0.8$ for remaining videos. It can be seen from Figure~\ref{fig:ablation_epoch_cmp} that \texttt{ILCAS} converges $1.15$--$3.22\times$ faster than \texttt{ILCAS} \textit{w/o Expert}. This is because \texttt{ILCAS} uses the IL framework to train the agent with the expert to guide the optimal direction of convergence. This advantage of IL is often referred to as \textit{high sample efficiency}, which has been evidenced by previous works~\cite{huang2020quality}\cite{li2022apprenticeship}. Finally, it is also interesting to find that \texttt{ILCAS} converges $1.50$--$3.16\times$ faster than \texttt{ILCAS} \textit{w/o MF}. One possible reason is that it is more efficient for \texttt{ILCAS} to mine the behavior pattern of expert with motion feature maps. }

\begin{figure}[t]
    \centering
    \includegraphics[width=0.4\textwidth]{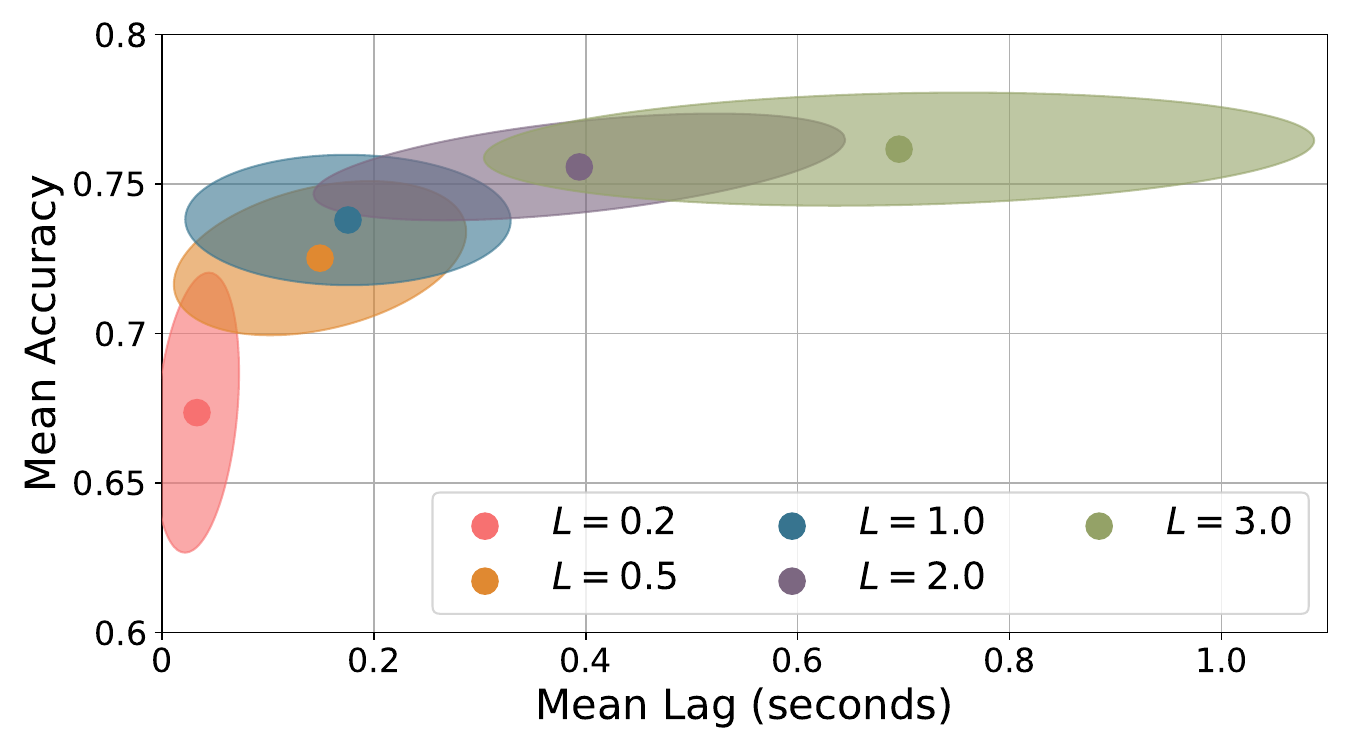}
    \vspace{-0.2cm}
    \caption{The mean lag vs. mean accuracy of \texttt{ILCAS} with different $L$. The $1$-$\sigma$ ellipses mark the performance variance of each solution. The center dots in each ellipse indicate the average values of all testing results of each solution.}
    \label{fig:ablation_param_L}
\end{figure}

\noindent \textbf{Impacts of parameter $L$.} Figure~\ref{fig:ablation_param_L} reports the performance of \texttt{ILCAS} on \textit{Dash1} video under different values of $L$.  Overall, increasing $L$ will improve the accuracy of \texttt{ILCAS}, especially when $L$ is relatively low (e.g., $L \le 1.0$). However, such performance gain will gradually diminish when $L$ is sufficiently high (e.g., $L>1.0$). Besides, with the increase of $L$, the mean upload lag of \texttt{ILCAS} becomes higher and the lag variance of \texttt{ILCAS} also becomes larger. This suggests that in practice, it is unnecessary to set $L$ too high for \texttt{ILCAS}, as it will not significantly improve the task accuracy but increase the upload lag and lag variance. 

\begin{figure}[t]
    \centering
    \subfigure[Mean accuracy]{
    \includegraphics[width=0.4\textwidth]{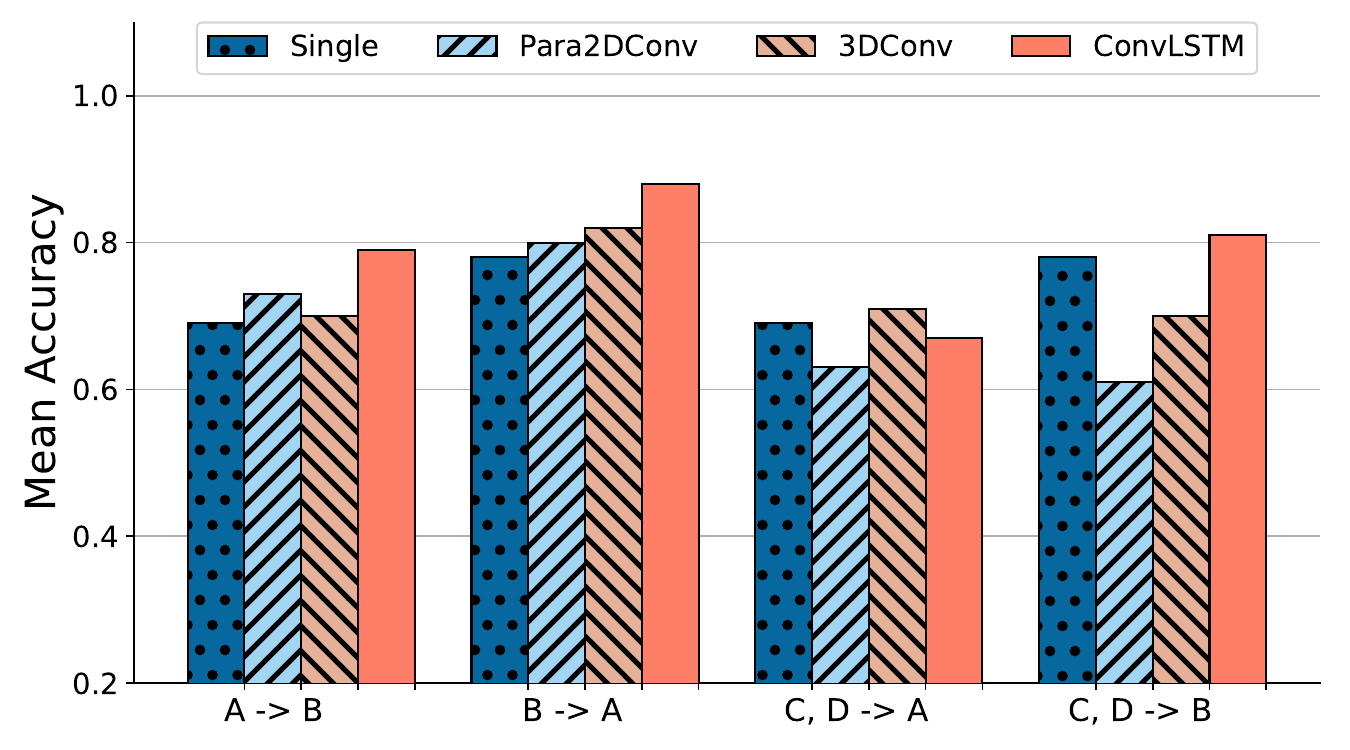}
    \label{subfig:cross_mean_acc}
    \vspace{-0.5cm}
    }
    \hspace{0.5cm}
    \subfigure[Mean lag]{
    \includegraphics[width=0.4\textwidth]{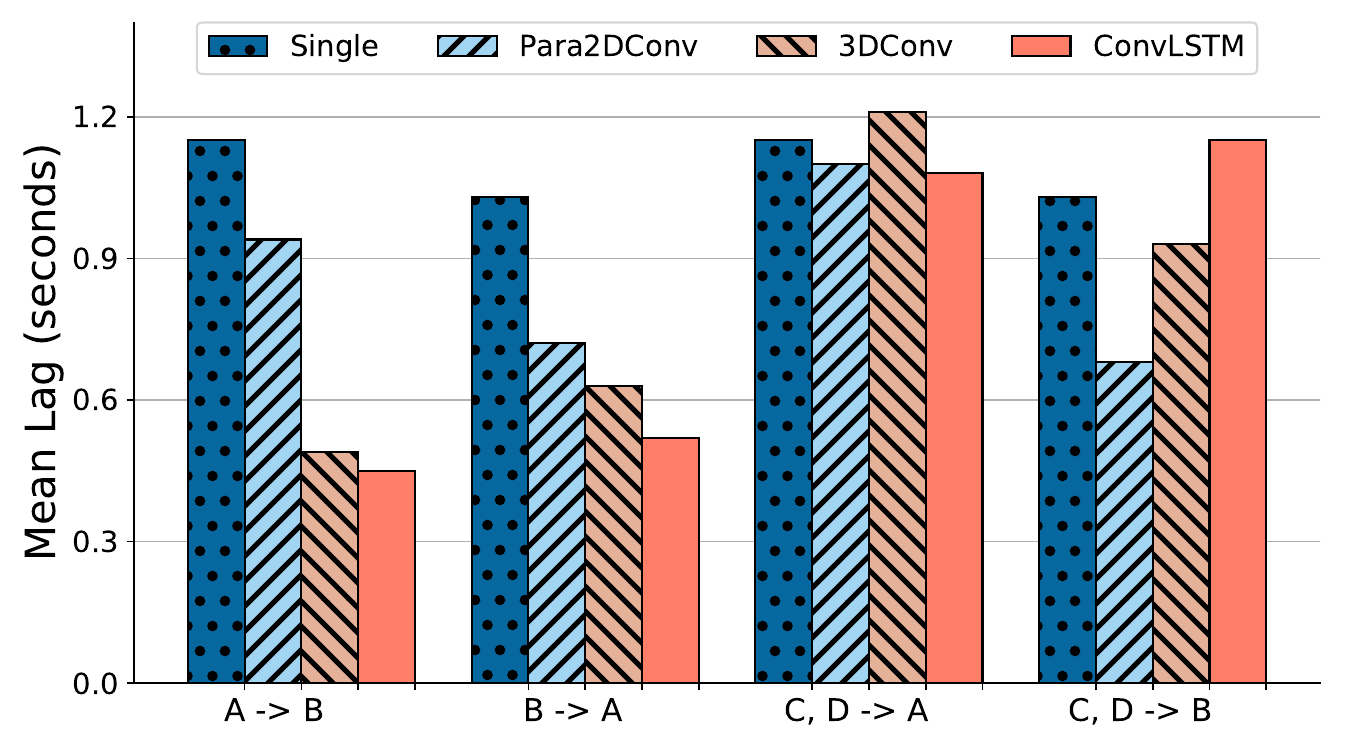}
    \label{subfig:cross_mean_lag}
    \vspace{-0.5cm}
    }
    \caption{The performance of different methods on \textit{AI City} dataset. X $\rightarrow$ Y means X is source camera and Y is target camera.}
    \label{fig:cross_mean_acc_lag_methods}
\end{figure}

\subsection{Effectiveness of Cross-Camera Collaboration Scheme}

In this part, we conduct experiments to verify the effectiveness of our proposed cross-camera collaboration scheme.
We set the correlation threshold $\mathbb{S}_{thresh} = 70\%$ and $\mathbb{T}_{thresh}=90\%$. 
We compare the performance of \texttt{ILCAS} with and without the collaboration scheme on \textit{AI City} dataset, denoted as \textit{Single} and \textit{ConvLSTM}, respectively. Additionally, we also design the following methods to process the motion feature maps from spatio-temporal correlated cameras for comparison:
\begin{itemize}
    \item \textit{Para2DConv}: This approach uses parallel 2D CNN sub-networks (described in Figure~\ref{subfig:cnn_subnet}) instead of ConvLSTM model to process the motion feature maps from correlated cameras, then merge their outputs and feed them into the FC layer.
    \item \textit{3DConv}~\cite{3dconv}: This approach replaces the ConvLSTM model with 3D CNN to extract hidden features from motion feature maps.
\end{itemize}

As shown in Figure~\ref{fig:cross_mean_acc_lag_methods} group 1, we first take camera B as the target camera and camera A as the source camera. It can be seen that each cross-camera-enhanced methods achieves better performance than \textit{Single}. Even if the $\textit{Para2DConv}$ uses the correlated feature maps as inputs straightforwardly, it still achieves $4\%$ accuracy improvement and reduces the mean lag by $22.3\%$.
By using 3D CNN to extract spatio-temporal information, the $\textit{3DConv}$ improves the mean accuracy by $1\%$ and reduces the mean lag by $57.4\%$, compare to the \textit{Single}. 
Since the ConvLSTM network shows its superiority in utilizing the spatio-temporal correlations and predicting future content dynamics, our approach $\textit{ConvLSTM}$ outperforms every method by $6$-$10\%$ accuracy improvement and $10.2$-$60.9\%$ lag reduction.

We further analyze the influence of correlations between cameras. Take camera A as the target camera and camera B as the source camera in turn, as shown in group 2 of Figure~\ref{fig:cross_mean_acc_lag_methods}, the performance of camera A is improved by $2$-$10\%$ and $43.1$-$49.5\%$ in terms of accuracy and reduction of lag, respectively.
Both camera A and camera B can benefit from one another due to their high spatio-temporal correlations, which indicates that the potential for cross-camera enhancement universally exists in highly-correlated cameras.
We next introduce cameras C and D as source cameras, which have weak correlations to target cameras A and B. As shown in group 3 and group 4, the cross-camera scheme does not show significant improvement compared to the \textit{Single} approach. 
One possible reason is that the motion feature maps from the weakly-correlated cameras C and D may contain some out-of-stale information that misleads the agent, leading to performance degradation. 
This highlights the importance of analyzing the camera correlations in cross-camera-based configuration adaption.

\section{Related Work}
\label{sec:related_work}
\noindent \textbf{Streaming configuration optimization for live VA.} Due to the limited computing power of front-end devices, streaming the camera videos over the network to resource-abundant edge/cloud servers for VA processing is prevalent\Revision{~\cite{zhang2023owl}\cite{yuan2023accdecoder}\cite{lei2022batch}}. In this context, optimizing the streaming configurations to balance the task accuracy and resource consumption has been extensively studied~\cite{zhang2022casva}. For instance, AWStream~\cite{zhang2018awstream} identifies the profile that characterizes the accuracy and bandwidth trade-off of different configurations, and adjusts the data rates based on such profile to match the available bandwidth while maximizing the application accuracy. 
CASVA~\cite{zhang2022casva} is a RL-based streaming system that aims to maximize the task accuracy and minimize the upload lag for each chunk by configuration selection. \texttt{ILCAS} differs from existing works in two aspects. First, \texttt{ILCAS} implements the state-of-the-art imitation learning framework GAIL~\cite{generative2016ho} to improve its efficiency in exploring the configuration space. Second, \texttt{ILCAS} models content dynamics as motion feature map, which allows \texttt{ILCAS} to select configurations that match the video contents.

\noindent \textbf{Imitation learning in networking.} Recent years have witnessed the successful applications of imitation learning (IL) in networking fields, such as task scheduling~\cite{wang2022imitation}, computation offloading~\cite{yu2020intelligent}, and adaptive bitrate streaming~\cite{huang2020quality}. Wang et al.~\cite{wang2022imitation} propose an IL-enabled online task scheduling algorithm for vehicular edge computing (VEC) networks to minimize the system energy consumption under the task latency constraints. Yu et al.~\cite{yu2020intelligent} formulate the computation offloading problem in mobile edge computing (MEC) networks as the multi-label classification problem, and design an efficient model to solve the problem based on behavioral cloning. 
In adaptive bitrate streaming, Comyco~\cite{huang2020quality} trains a policy network via IL to select bitrates that maximize the perceptual video quality for users. 
Unlike these systems, to the best of our knowledge, \texttt{ILCAS} is the first VA streaming system that exploits IL for configuration optimization while adapting to the network environment and video content changes.

\noindent \textbf{Cross camera collaboration.} Video cameras are ubiquitous in modern cities and cameras closely located in the same region (e.g., traffic corner) often share some spatial and temporal correlations~\cite{jain2019scaling}. Such correlations have been identified and explored to optimize the video analytics process. 
CrossRoI~\cite{guo2021crossroi} offline establishes cross-camera correlations and online filters out the redundant information in multiple camera videos to reduce the redundant data transmitted to remote server for processing. 
Respire~\cite{dai2022respire} characterizes the spatio-temporal redundancy of video frames from multiple cameras, and only uploads the most informative frames to cloud servers for analytics to reduce transmission cost. 
\Revision{Polly~\cite{jingzong2023cross} shares the video analytics results across correlated cameras to reduce redundant analytic work.}
Different from these systems, \texttt{ILCAS} is specially designed for configuration optimization for live VA streaming. It for the first time exploits the camera correlations to achieve more adaptive and robust configuration selection.

\section{Conclusion}
\label{sec:conclusion}
In this paper, we present \texttt{ILCAS}, the first imitation learning-enabled configuration adaption system for live VA streaming. Unlike DRL-based solutions that heavily rely on reward functions to train the agent, \texttt{ILCAS} trains the agent with expert's demonstrations through the advanced Generative Adversarial Imitation Learning (GAIL) framework, which improves the overall performance of \texttt{ILCAS}. To provide demonstrations for training the agent, we design the expert as an efficient dynamic programming algorithm. Besides, \texttt{ILCAS} extracts motion feature map from camera codec to capture the video content dynamics for more adaptive configuration selection. Additionally, \texttt{ILCAS} also incorporates a cross-camera collaboration scheme to exploit the camera correlations for better performance in a camera network. Experiments demonstrate that \texttt{ILCAS} outperforms existing state-of-the-art solutions by improving the mean accuracy by 2.0-20.9\% and reducing the mean lag by 19.9-85.3\%. We also show that the cross-camera collaboration scheme improves the performance of \texttt{ILCAS} in a dense camera network.

\ifCLASSOPTIONcompsoc
  \section*{Acknowledgments}
\else
  \section*{Acknowledgment}
\fi
The work was supported in part by the Basic Research Project No. HZQB-KCZYZ-2021067 of Hetao Shenzhen-HK S\&T Cooperation Zone, by NSFC (Grant No. 62293482 and No. 62102342), the Guangdong Basic and Applied Basic Research Foundation (Grant No. 2023A1515012668), the Shenzhen Science and Technology Program (Grant No. RCBS20221008093120047), the Shenzhen Outstanding Talents Training Fund 202002, the Guangdong Research Projects No. 2017ZT07X152 and No. 2019CX01X104, the Young Elite Scientists Sponsorship Program of CAST (Grant No. 2022QNRC001), the Guangdong Provincial Key Laboratory of Future Networks of Intelligence (Grant No. 2022B1212010001), the Shenzhen Key Laboratory of Big Data and Artificial Intelligence (Grant No. ZDSYS201707251409055). 

\ifCLASSOPTIONcaptionsoff
  \newpage
\fi

\bibliographystyle{IEEEtran}
\bibliography{IEEEabrv,references}

\begin{thebibliography}{10}
\providecommand{\url}[1]{#1}
\csname url@samestyle\endcsname
\providecommand{\newblock}{\relax}
\providecommand{\bibinfo}[2]{#2}
\providecommand{\BIBentrySTDinterwordspacing}{\spaceskip=0pt\relax}
\providecommand{\BIBentryALTinterwordstretchfactor}{4}
\providecommand{\BIBentryALTinterwordspacing}{\spaceskip=\fontdimen2\font plus
\BIBentryALTinterwordstretchfactor\fontdimen3\font minus
  \fontdimen4\font\relax}
\providecommand{\BIBforeignlanguage}[2]{{%
\expandafter\ifx\csname l@#1\endcsname\relax
\typeout{** WARNING: IEEEtran.bst: No hyphenation pattern has been}%
\typeout{** loaded for the language `#1'. Using the pattern for}%
\typeout{** the default language instead.}%
\else
\language=\csname l@#1\endcsname
\fi
#2}}
\providecommand{\BIBdecl}{\relax}
\BIBdecl

\bibitem{abdullah2014traffic}
T.~Abdullah, A.~Anjum, M.~F. Tariq, Y.~Baltaci, and N.~Antonopoulos, ``Traffic
  monitoring using video analytics in clouds,'' in \emph{IEEE/ACM 7th
  International Conference on Utility and Cloud Computing}, 2014, pp. 39--48.

\bibitem{guo2022minimizing}
Q.~Guo, J.~Peng, W.~Xu, W.~Liang, X.~Jia, Z.~Xu, Y.~Yang, and M.~Wang,
  ``{Minimizing the longest tour time among a fleet of UAVs for disaster area
  surveillance},'' \emph{IEEE Transactions on Mobile Computing}, vol.~21,
  no.~7, pp. 2451--2465, 2022.

\bibitem{ren2015faster}
S.~Ren, K.~He, R.~Girshick, and J.~Sun, ``{Faster R-CNN: Towards real-time
  object detection with region proposal networks},'' \emph{{Advances in Neural
  Information Processing Systems}}, vol.~28, 2015.

\bibitem{zhao2018icnet}
H.~Zhao, X.~Qi, X.~Shen, J.~Shi, and J.~Jia, ``{ICNet for real-time semantic
  segmentation on high-resolution images},'' in \emph{Proceedings of the ECCV},
  2018, pp. 405--420.

\bibitem{miao2023omnisense}
M.~Zhang, Y.~Zhu, L.~Shen, F.~Wang, and J.~Liu, ``Omnisense: Towards
  edge-assisted online analytics for 360-degree videos,'' in \emph{Proceedings
  of the IEEE INFOCOM}, 2023, pp. 1--10.

\bibitem{zhang2018awstream}
B.~Zhang, X.~Jin, S.~Ratnasamy, J.~Wawrzynek, and E.~A. Lee, ``{AWStream:
  Adaptive wide-area streaming analytics},'' in \emph{Proceedings of the ACM
  SIGCOMM}, 2018, pp. 236--252.

\bibitem{zhang2017live}
H.~Zhang, G.~Ananthanarayanan, P.~Bodik, M.~Philipose, P.~Bahl, and M.~J.
  Freedman, ``Live video analytics at scale with approximation and
  delay-tolerance,'' in \emph{USENIX NSDI}, 2017, pp. 377--392.

\bibitem{jiang2018chameleon}
J.~Jiang, G.~Ananthanarayanan, P.~Bodik, S.~Sen, and I.~Stoica, ``{Chameleon:
  Scalable adaptation of video analytics},'' in \emph{Proceedings of the ACM
  SIGCOMM}, 2018, pp. 253--266.

\bibitem{hsieh2018focus}
K.~Hsieh, G.~Ananthanarayanan, P.~Bodik, S.~Venkataraman, P.~Bahl,
  M.~Philipose, P.~B. Gibbons, and O.~Mutlu, ``{Focus: Querying large video
  datasets with low latency and low cost},'' in \emph{USENIX OSDI}, 2018, pp.
  269--286.

\bibitem{zhang2022casva}
M.~Zhang, F.~Wang, and J.~Liu, ``{CASVA: Configuration-adaptive streaming for
  live video analytics},'' in \emph{Proceedings of the IEEE INFOCOM}, 2022, pp.
  2168--2177.

\bibitem{zhang2022maxim}
R.~Zhang, Y.~Zhou, F.~Wang, and Z.~Wang, ``{Maxim: DRL-based cross-camera
  streaming configuration for real-time video analytics},'' in \emph{IEEE
  ICME}, 2022, pp. 1--6.

\bibitem{hussein2017imitation}
A.~Hussein, M.~M. Gaber, E.~Elyan, and C.~Jayne, ``{Imitation learning: A
  survey of learning methods},'' \emph{ACM Computing Surveys}, vol.~50, no.~2,
  pp. 1--35, 2017.

\bibitem{Zhang_2016_CVPR}
B.~Zhang, L.~Wang, Z.~Wang, Y.~Qiao, and H.~Wang, ``{Real-time action
  recognition with enhanced motion vector CNNs},'' in \emph{Proceedings of the
  IEEE CVPR}, 2016.

\bibitem{le2022survey}
L.~Le~Mero, D.~Yi, M.~Dianati, and A.~Mouzakitis, ``A survey on imitation
  learning techniques for end-to-end autonomous vehicles,'' \emph{IEEE
  Transactions on Intelligent Transportation Systems}, vol.~23, no.~9, pp.
  14\,128--14\,147, 2022.

\bibitem{sutton2018reinforcement}
R.~S. Sutton and A.~G. Barto, \emph{Reinforcement learning: An
  introduction}.\hskip 1em plus 0.5em minus 0.4em\relax MIT press, 2018.

\bibitem{cross}
\BIBentryALTinterwordspacing
``{2022 AI city challenge - track 1: City-scale multi-camera vehicle
  tracking}.'' [Online]. Available:
  \url{https://www.aicitychallenge.org/2022-data-and-evaluation/}
\BIBentrySTDinterwordspacing

\bibitem{redmon2016you}
J.~Redmon, S.~Divvala, R.~Girshick, and A.~Farhadi, ``{You only look once:
  Unified, real-time object detection},'' in \emph{Proceedings of the IEEE
  CVPR}, 2016, pp. 779--788.

\bibitem{ConvLSTM}
X.~Shi, Z.~Chen, H.~Wang, D.-Y. Yeung, W.-K. Wong, and W.-c. Woo,
  ``{Convolutional LSTM network: A machine learning approach for precipitation
  nowcasting},'' \emph{{Advances in Neural Information Processing Systems}},
  vol.~28, 2015.

\bibitem{generative2016ho}
J.~Ho and S.~Ermon, ``Generative adversarial imitation learning,'' in
  \emph{Advances in Neural Information Processing Systems}, vol.~29, 2016.

\bibitem{schulman2017proximal}
J.~Schulman, F.~Wolski, P.~Dhariwal, A.~Radford, and O.~Klimov, ``Proximal
  policy optimization algorithms,'' \emph{arXiv preprint arXiv:1707.06347},
  2017.

\bibitem{sr520bridgetolling}
\BIBentryALTinterwordspacing
``{SR 520 bridge tolling}.'' [Online]. Available:
  \url{https://www.wsdot.wa.gov/Tolling/520/default.htm}
\BIBentrySTDinterwordspacing

\bibitem{jain2020spatula}
S.~Jain, X.~Zhang, Y.~Zhou, G.~Ananthanarayanan, J.~Jiang, Y.~Shu, P.~Bahl, and
  J.~Gonzalez, ``{Spatula: Efficient cross-camera video analytics on large
  camera networks},'' in \emph{2020 IEEE/ACM Symposium on Edge Computing
  (SEC)}, 2020, pp. 110--124.

\bibitem{dash1}
\BIBentryALTinterwordspacing
``{Skyline views - Chicago 4K - driving downtown}.'' [Online]. Available:
  \url{https://www.youtube.com/watch?v=O-YPkapBT8g\&t=77s}
\BIBentrySTDinterwordspacing

\bibitem{dash2}
\BIBentryALTinterwordspacing
``{London 4K - night drive - UK}.'' [Online]. Available:
  \url{https://www.youtube.com/watch?v=Ujyu8foke60}
\BIBentrySTDinterwordspacing

\bibitem{stationary1}
\BIBentryALTinterwordspacing
``{4K road traffic video for object detection and tracking}.'' [Online].
  Available: \url{https://www.youtube.com/watch?v=MNn9qKG2UFI\&t=81s}
\BIBentrySTDinterwordspacing

\bibitem{stationary2}
\BIBentryALTinterwordspacing
``{4K video of highway traffic}.'' [Online]. Available:
  \url{https://www.youtube.com/watch?v=KBsqQez-O4w}
\BIBentrySTDinterwordspacing

\bibitem{stationary3}
\BIBentryALTinterwordspacing
``{Road traffic video for object recognition}.'' [Online]. Available:
  \url{https://www.youtube.com/watch?v=wqctLW0Hb\_0}
\BIBentrySTDinterwordspacing

\bibitem{van2016http}
J.~van~der Hooft, S.~Petrangeli, T.~Wauters, R.~Huysegems, P.~R. Alface,
  T.~Bostoen, and F.~De~Turck, ``{HTTP/2-based adaptive streaming of HEVC video
  over 4G/LTE networks},'' \emph{IEEE Communications Letters}, vol.~20, no.~11,
  pp. 2177--2180, 2016.

\bibitem{huang2020quality}
T.~Huang, C.~Zhou, X.~Yao, R.-X. Zhang, C.~Wu, B.~Yu, and L.~Sun,
  ``Quality-aware neural adaptive video streaming with lifelong imitation
  learning,'' \emph{IEEE Journal on Selected Areas in Communications}, vol.~38,
  no.~10, pp. 2324--2342, 2020.

\bibitem{li2022apprenticeship}
W.~Li, J.~Huang, S.~Wang, C.~Wu, S.~Liu, and J.~Wang, ``An apprenticeship
  learning approach for adaptive video streaming based on chunk quality and
  user preference,'' \emph{IEEE Transactions on Multimedia}, 2022.

\bibitem{3dconv}
D.~Tran, L.~Bourdev, R.~Fergus, L.~Torresani, and M.~Paluri, ``{Learning
  spatiotemporal features with 3D convolutional networks},'' in
  \emph{Proceedings of the IEEE ICCV}, 2015, pp. 4489--4497.

\bibitem{zhang2023owl}
R.-X. Zhang, C.~Li, C.~Wu, T.~Huang, and L.~Sun, ``{Owl: A pre-and
  post-processing framework for video analytics in low-light surroundings},''
  in \emph{Proceedings of the IEEE INFOCOM}, 2023.

\bibitem{yuan2023accdecoder}
T.~Yuan, L.~Mi, W.~Wang, H.~Dai, and X.~Fu, ``{AccDecoder: Accelerated decoding
  for neural-enhanced video analytics},'' in \emph{Proceedings of the IEEE
  INFOCOM}, 2023.

\bibitem{lei2022batch}
L.~Zhang, Y.~Zhang, X.~Wu, F.~Wang, L.~Cui, Z.~Wang, and J.~Liu, ``Batch
  adaptative streaming for video analytics,'' in \emph{Proceedings of the IEEE
  INFOCOM}, 2022, pp. 2158--2167.

\bibitem{wang2022imitation}
X.~Wang, Z.~Ning, S.~Guo, and L.~Wang, ``Imitation learning enabled task
  scheduling for online vehicular edge computing,'' \emph{IEEE Transactions on
  Mobile Computing}, vol.~21, no.~2, pp. 598--611, 2022.

\bibitem{yu2020intelligent}
S.~Yu, X.~Chen, L.~Yang, D.~Wu, M.~Bennis, and J.~Zhang, ``{Intelligent edge:
  Leveraging deep imitation learning for mobile edge computation offloading},''
  \emph{IEEE Wireless Communications}, vol.~27, no.~1, pp. 92--99, 2020.

\bibitem{jain2019scaling}
S.~Jain, G.~Ananthanarayanan, J.~Jiang, Y.~Shu, and J.~Gonzalez, ``Scaling
  video analytics systems to large camera deployments,'' in \emph{ACM
  HotMobile}, 2019, pp. 9--14.

\bibitem{guo2021crossroi}
H.~Guo, S.~Yao, Z.~Yang, Q.~Zhou, and K.~Nahrstedt, ``{CrossRoI: cross-camera
  region of interest optimization for efficient real time video analytics at
  scale},'' in \emph{Proceedings of the 12th ACM Multimedia Systems Conference
  (MMSys)}, 2021, pp. 186--199.

\bibitem{dai2022respire}
X.~Dai, P.~Yang, X.~Zhang, Z.~Dai, and L.~Yu, ``Respire: Reducing
  spatial–temporal redundancy for efficient edge-based industrial video
  analytics,'' \emph{IEEE Transactions on Industrial Informatics}, vol.~18,
  no.~12, pp. 9324--9334, 2022.

\bibitem{jingzong2023cross}
L.~Jingzong, L.~Liu, H.~Xu, S.~Wu, and C.~J. Xue, ``Cross-camera inference on
  the constrained edge,'' in \emph{Proceedings of the IEEE INFOCOM}, 2023.

\end{thebibliography}
\begin{IEEEbiography}[{\includegraphics[width=1in,height=1.25in,clip,keepaspectratio]{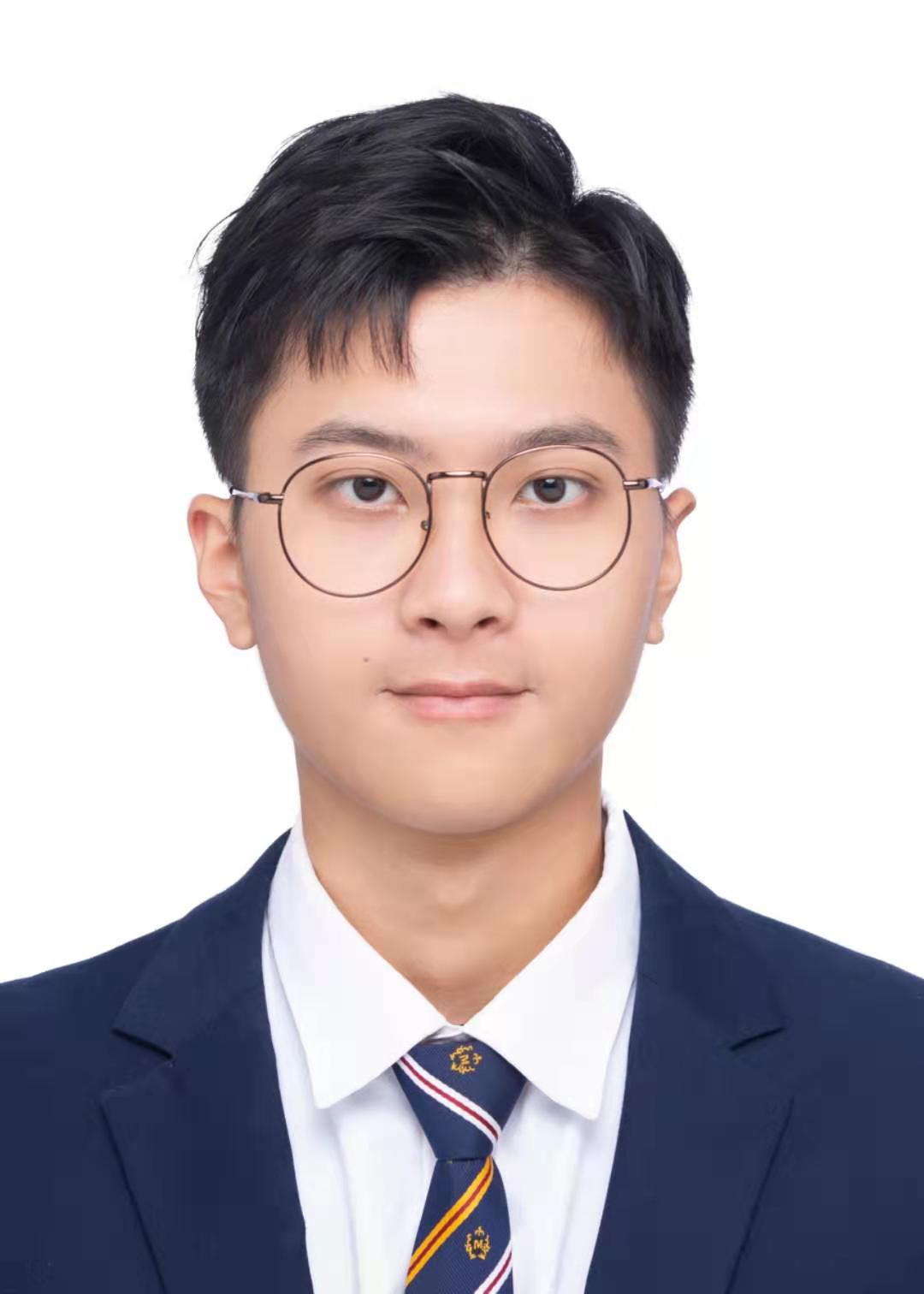}}]{Duo Wu} received the B.Eng. degree from the College of Information Science and Technology, Jinan University in 2022. He is currently pursuing the M.Phil. degree in the School of Science and Engineering, The Chinese University of Hong Kong, Shenzhen. His current research interests include multimedia networking, video analytics, video streaming and machine learning.
\end{IEEEbiography}

\begin{IEEEbiography}[{\includegraphics[width=1in,height=1.25in,clip,keepaspectratio]{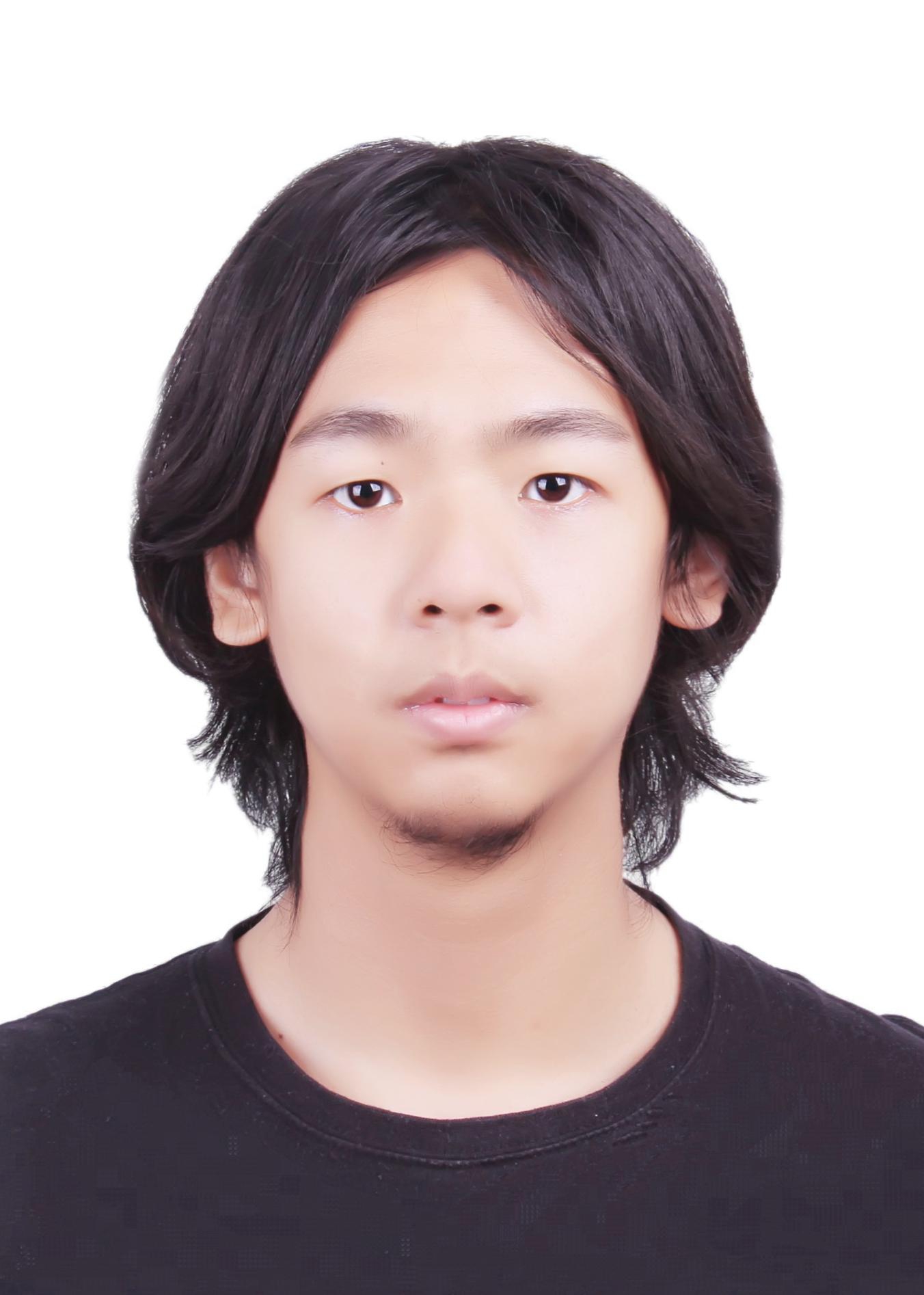}}]{Dayou Zhang} received the B.S. degree from the School of Information and Electronics, Beijing Institute of Technology, Beijing, China, in 2017. He is currently pursuing the Ph.D. degree in the School of Science and Engineering, The Chinese University of Hong Kong, Shenzhen, China. His areas of interest are multimedia networking, real-time video streaming, neural video coding, and machine learning.
\end{IEEEbiography}
\vspace{-1.5cm}
\begin{IEEEbiography}[{\includegraphics[width=1in,height=1.25in,clip,keepaspectratio]{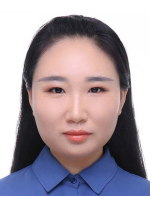}}]{Miao Zhang}
 (Student Member, IEEE) received the B.Eng. degree from Sichuan University in 2015 and the M.Eng. degree from Tsinghua University
in 2018. She is currently pursuing the Ph.D. degree with Simon Fraser University, Canada. Her research areas include cloud and edge computing and multimedia systems and applications.
\end{IEEEbiography}

\begin{IEEEbiography}[{\includegraphics[width=1in,height=1.25in,clip,keepaspectratio]{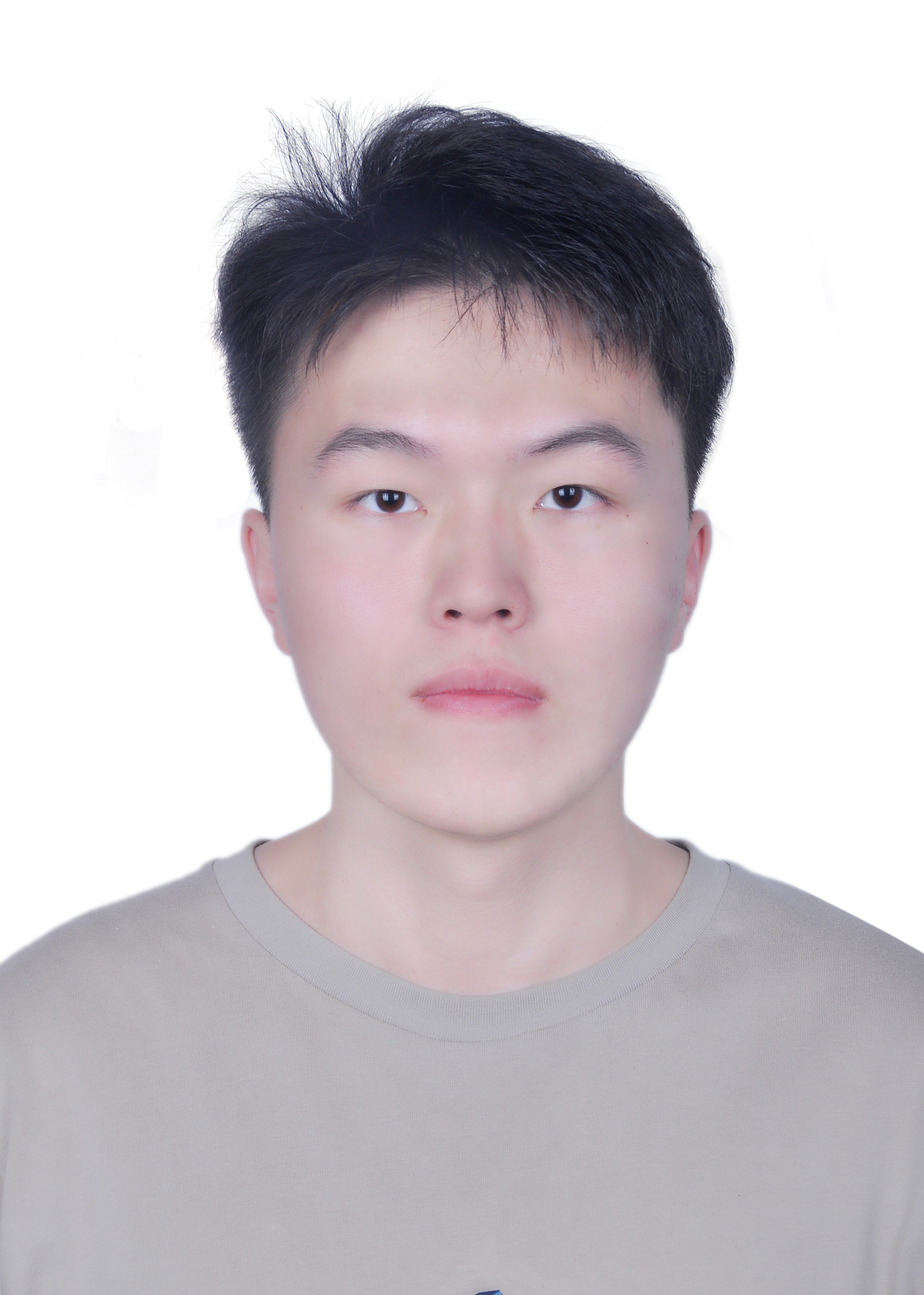}}]{Ruoyu Zhang} received the B.Eng. degree from the Zhuoyue Honors College, Hangzhou Danzi University in 2021. He is currently pursuing the M.Phil. degree in the School of Science and Engineering, The Chinese University of Hong Kong, Shenzhen. His research interests include multimedia networking, video analytics and machine learning.
\end{IEEEbiography}

\begin{IEEEbiography}[{\includegraphics[width=1in,height=1.25in,clip,keepaspectratio]{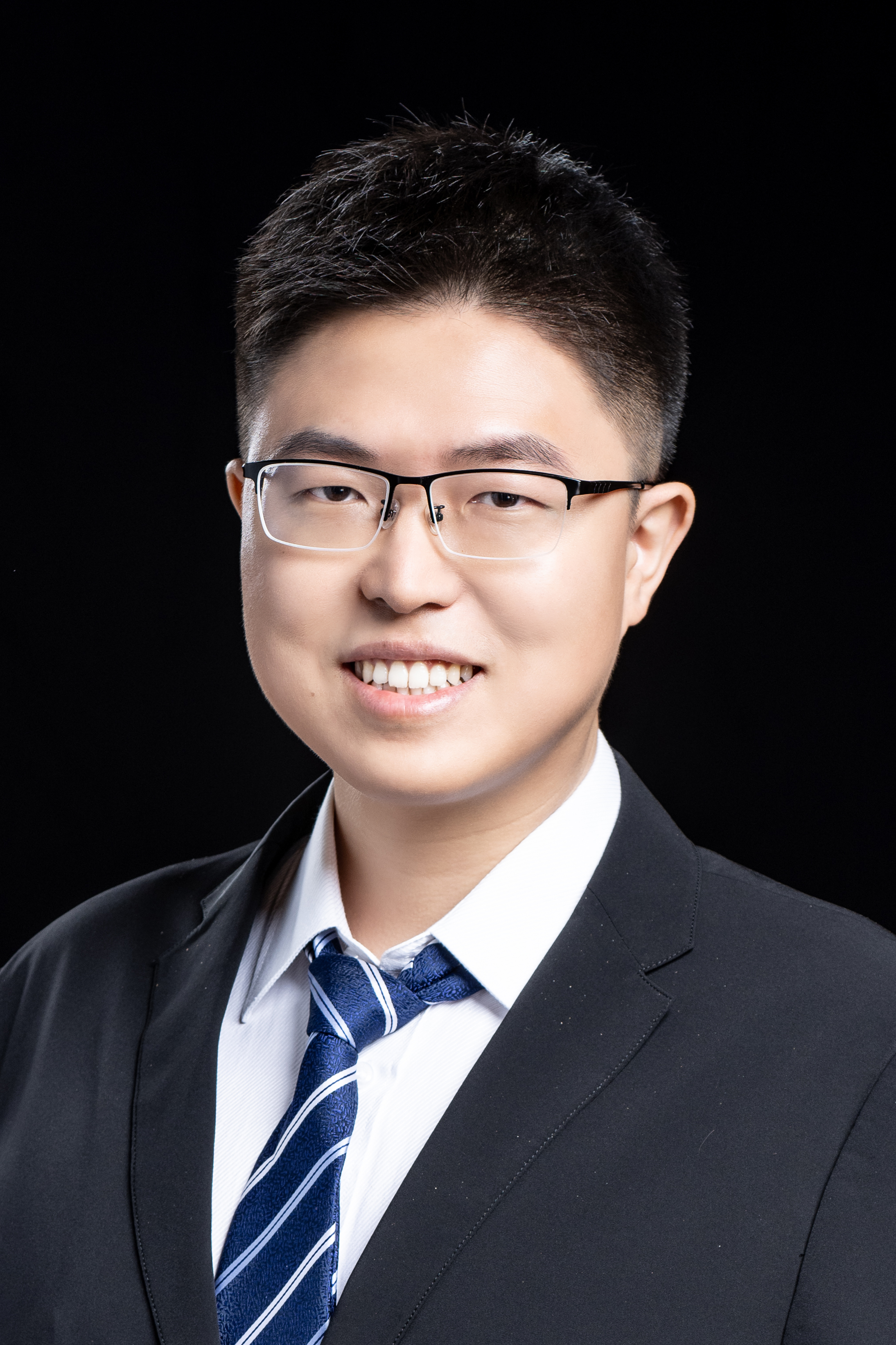}}]{Fangxin Wang}(S'15-M'20) is an assistant professor at The Chinese University of Hong Kong, Shenzhen (CUHKSZ). He received the Ph.D., M.Eng., and B.Eng. degree all in Computer Science and Technology from Simon Fraser University, Tsinghua University, and Beijing University of Posts and Telecommunications, respectively. 
Dr. Wang's research interests include Multimedia Systems and Applications, Cloud and Edge Computing, Deep Learning and Big Data Analytics, Distributed Networking and System. 
He has published more than 30 papers at top journal and conference papers, including INFOCOM, Multimedia, ToN, TMC, IOTJ, etc. 
\end{IEEEbiography}

\begin{IEEEbiography}[{\includegraphics[width=1in,height=1.25in,clip,keepaspectratio]{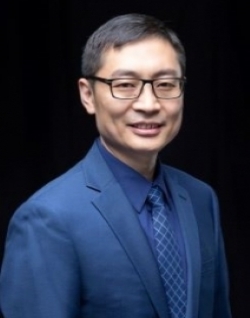}}]{Shuguang Cui}(S’99-M’05-SM’12-F’14)  received his Ph.D in Electrical Engineering from Stanford University, California, USA, in 2005. Afterwards, he has been working as assistant, associate, full, Chair Professor in Electrical and Computer Engineering at the Univ. of Arizona, Texas A\&M University, UC Davis, and CUHK at Shenzhen, respectively. He was selected as the Thomson Reuters Highly Cited Researcher and listed in the Worlds Most Influential Scientific Minds by ScienceWatch in 2014. 
He has also been serving as the area editor for IEEE Signal Processing Magazine, and associate editors for IEEE Transactions on Big Data, IEEE Transactions on Signal Processing. He is a member of the Steering Committee for IEEE Transactions on Big Data and the Chair of the Steering Committee for IEEE Transactions on Cognitive Communications and Networking. He was elected as an IEEE Fellow in 2013, an IEEE ComSoc Distinguished Lecturer in 2014, and IEEE VT Society Distinguished Lecturer in 2019. 
\end{IEEEbiography}
\end{document}